\documentclass[prd,preprint,superscriptaddress,tightenlines,nofootinbib,
  eqsecnum,showpacs]{revtex4}

\usepackage{amsmath}
\usepackage{amsfonts}
\usepackage{amssymb}
\usepackage{bm}
\usepackage{hyperref}
\usepackage{mathrsfs}
\usepackage{graphicx}

\usepackage{ulem}
\usepackage[usenames]{color}

\definecolor{darkgreen}{rgb}{0,0.5,0}

\hypersetup{
    bookmarks=true,         % show bookmarks bar?
    unicode=false,          % non-Latin characters in Acrobat???s bookmarks
    pdftoolbar=true,        % show Acrobat???s toolbar?
    pdfmenubar=true,        % show Acrobat???s menu?
    pdffitwindow=false,     % window fit to page when opened
    pdfstartview={FitH},    % fits the width of the page to the window
    pdftitle={My title},    % title
    pdfauthor={Author},     % author
    pdfsubject={Subject},   % subject of the document
    pdfcreator={Creator},   % creator of the document
    pdfproducer={Producer}, % producer of the document
    pdfkeywords={keyword1} {key2} {key3}, % list of keywords
    pdfnewwindow=true,      % links in new window
    colorlinks=true,       % false: boxed links; true: colored links
    linkcolor=red,          % color of internal links
    citecolor=cyan,        % color of links to bibliography
    filecolor=magenta,      % color of file links
    urlcolor=darkgreen,           % color of external links
    linktocpage=true
}

\allowdisplaybreaks
\DeclareSymbolFontAlphabet{\mathrsfs}{rsfs}
\DeclareMathAlphabet{\mathcal}{OMS}{cmsy}{m}{n}

\newcommand{\ud}{\mathrm{d}}
\newcommand{\ui}{\mathrm{i}} 
\newcommand{\ue}{\mathrm{e}} 
\newcommand{\beq}{\begin{equation}}
\newcommand{\eeq}{\end{equation}}

\newcounter{theorem} 

\begin{document}

\title{Energy and periastron advance of compact binaries \\on circular
  orbits at the fourth post-Newtonian order}

\author{Laura Bernard}\email{laura.bernard@tecnico.ulisboa.pt}
\affiliation{$\mathcal{G}\mathbb{R}\varepsilon{\mathbb{C}}\mathcal{O}$,
  Institut d'Astrophysique de Paris,\\ UMR 7095, CNRS, Sorbonne
  Universit{\'e}s \& UPMC Univ Paris 6,\\ 98\textsuperscript{bis}
  boulevard Arago, 75014 Paris, France} \affiliation{CENTRA,
  Departamento de F\'{\i}sica, Instituto Superior T{\'e}cnico -- IST,
  Universidade de Lisboa -- UL, Avenida Rovisco Pais 1, 1049 Lisboa,
  Portugal}

\author{Luc Blanchet}\email{blanchet@iap.fr}
\affiliation{$\mathcal{G}\mathbb{R}\varepsilon{\mathbb{C}}\mathcal{O}$,
  Institut d'Astrophysique de Paris,\\ UMR 7095, CNRS, Sorbonne
  Universit{\'e}s \& UPMC Univ Paris 6,\\ 98\textsuperscript{bis}
  boulevard Arago, 75014 Paris, France}

\author{Alejandro Boh\'e}\email{alejandro.bohe@aei.mpg.de}
\affiliation{Albert Einstein Institut, Am Muehlenberg 1, 14476
  Potsdam-Golm, Germany}

\author{Guillaume Faye}\email{faye@iap.fr}
\affiliation{$\mathcal{G}\mathbb{R}\varepsilon{\mathbb{C}}\mathcal{O}$,
  Institut d'Astrophysique de Paris,\\ UMR 7095, CNRS, Sorbonne
  Universit{\'e}s \& UPMC Univ Paris 6,\\ 98\textsuperscript{bis}
  boulevard Arago, 75014 Paris, France}

\author{Sylvain Marsat}\email{sylvain.marsat@aei.mpg.de}
\affiliation{Albert Einstein Institut, Am Muehlenberg 1, 14476
  Potsdam-Golm, Germany} 

\date{\today}

\begin{abstract}
  In this paper, we revisit and complete our preceding work on the
  Fokker Lagrangian describing the dynamics of compact binary systems
  at the fourth post-Newtonian (4PN) order in harmonic coordinates. We
  clarify the impact of the non-local character of the Fokker
  Lagrangian or the associated Hamiltonian on both the conserved
  energy and the relativistic periastron precession for circular
  orbits. We show that the non-locality of the action, due to the
  presence of the tail effect at the 4PN order, gives rise to an extra
  contribution to the conserved integral of energy with respect to the
  Hamiltonian computed on shell, which was not taken into account in
  our previous work. We also provide a direct derivation of the
  periastron advance by taking carefully into account this
  non-locality. We then argue that the infra-red (IR) divergences in
  the calculation of the gravitational part of the action are
  problematic, which motivates us to introduce a second ambiguity
  parameter, in addition to the one already assumed previously. After
  fixing these two ambiguity parameters by requiring that the
  conserved energy and the relativistic periastron precession for
  circular orbits are in agreement with numerical and analytical
  gravitational self-force calculations, valid in the limiting case of
  small mass ratio, we find that our resulting Lagrangian is
  physically equivalent to the one obtained in the ADM Hamiltonian
  approach.
\end{abstract}


\maketitle

\section{Introduction} 
\label{sec:intro}

This paper is a follow up to Ref.~\cite{BBBFM16a}, in which we derived
the equations of motion of compact binary systems (without spins) at
the 4PN approximation of general relativity.\footnote{We shall refer
  to Ref.~\cite{BBBFM16a} as ``Paper~I'' henceforth. As usual the
  $n$PN order corresponds to post-Newtonian corrections up to order
  $(v/c)^{2n}$ beyond the Newtonian acceleration in the equations of
  motion.} In Paper~I, we gave a short account of numerous past works
on the PN equations of motion of compact binaries
(see~\cite{Bliving14} for a more exhaustive review). At the 3PN order,
the equations were independently derived using different methods:
Hamiltonian
formalism~\cite{JaraS98,JaraS99,JaraS00,DJSpoinc,DJSequiv,DJSdim},
harmonic coordinates
approach~\cite{BF00,BFeom,BFreg,ABF01,BI03CM,BDE04}, surface-integral
method~\cite{IFA00,IFA01,itoh1,itoh2,ItohLR}, and effective field
theory (EFT)~\cite{FS3PN}.

At the 4PN order, partial results were first obtained within the
Hamiltonian formalism in ADM-like
coordinates~\cite{JaraS12,JaraS13,JaraS15} and the EFT
approach~\cite{FS4PN}. The 4PN Hamiltonian was then completed in
Refs.~\cite{DJS14,DJS16} by adding a non-local (in time) contribution
related to gravitational-wave tails, known from Refs.~\cite{BD88,B93}
for general matter systems. An alternative derivation of the 4PN
dynamics, including the same non-local tail piece, was achieved in
Paper~I by computing the Fokker Lagrangian in harmonic
coordinates. This result agreed with the partial results of the
EFT~\cite{FS4PN}, and also with most of the terms in the 4PN
Hamiltonian~\cite{JaraS12,JaraS13,JaraS15,DJS14,DJS16}. However, it
disagreed with a few terms [see Eqs.~(5.18)--(5.19) in Paper~I]. Part
of the discrepancy was due to the fact that Refs.~\cite{DJS14,DJS16}
and Paper~I use different prescriptions to handle the non-local tail
term, as discussed in Sec.~VC of Paper~I and in more details below.

An important point of comparison for the PN results is given by
gravitational self-force (GSF) calculations, valid in the small mass
ratio limit (see Refs.~\cite{Detweilerorleans,Barackorleans} for
reviews). The recent years have seen a lot of progress on numerical or
analytical calculations of two gauge-invariant quantities, the
conserved energy and the advance of the periastron for circular
orbits, allowing for unambiguous comparisons with the small mass-ratio
limit of the PN results. The GSF energy for circular orbits has been
computed numerically in~\cite{BDLW10b,LBW12,LBB12} and analytically
(in the form of a PN expansion) in~\cite{BiniD13}, while the circular
orbit limit of the periastron advance has been obtained numerically
in~\cite{BDS10,Letal11,vdM16} and analytically
in~\cite{D10sf,DJS15eob,DJS16}. The GSF calculations of the energy and
periastron advance generally rely on the so-called first law of black
hole binaries~\cite{LBW12,L15}, which was initially derived at 3PN
order, but has recently been shown to be valid at the 4PN order,
taking into account the non-local tail term~\cite{BL16}. Both in
Paper~I and in Refs.~\cite{DJS14,DJS16}, a single free parameter was
introduced, associated to the 4PN tail term, and was fixed by
comparing the conserved energy for circular orbits with GSF
results~\cite{BDLW10b, LBW12, BiniD13}.

However, the treatment of the non-locality in the derivation of this
quantity from the action was different in the two approaches. In the
present paper, we investigate in more details the derivation of both
the energy and the periastron advance from the non-local action. With
respect to Paper~I, we show the appearance of a new term in the
conserved integral of energy that resolves our disagreement with
Ref.~\cite{DJS14} on this issue. This is achieved within our initial
approach, based on the direct use of the original action containing
the non-local 4PN tail contribution, \textit{i.e.}, not applying any
non-local shift to transform the non-local action into a local one as
advocated in Ref.~\cite{DJS14}. We thoroughly compute the new term
using Fourier series, as well as a similar term present in the
conserved integral of angular momentum. While the result for the
conserved energy is affected by the former term, the periastron
advance remains unchanged, as the extra contributions coming from the
energy and angular momentum cancel out in this quantity.

The computation in Paper~I was supplemented by a dimensional
regularization for treating the ultra-violet (UV) divergences
associated with point particles, and by a Hadamard regularization for
curing the infra-red (IR) divergences that occur at the bound at
infinity of integrals entering the gravitational (Einstein-Hilbert)
part of the action. Past experience on the UV regularization at the
3PN order, both for the equations of motion~\cite{BFeom} and the
radiation field~\cite{BIJ02}, shows that different implementations of
the Hadamard regularization can give different physical results, but
with differences that were limited to a small subset of terms. At the
time, this unsatisfactory situation led to the introduction of unknown
ambiguity parameters, that were determined by the consistent
replacement of the Hadamard regularization by the dimensional
regularization. In the present paper we argue that the IR
regularization of the bound at infinity is problematic, as different
prescriptions for that regularization may lead to different
results. However, we conjecture, based on some preliminary
work~\cite{BBBFM16c}, that the difference between different
prescriptions for the IR regularization is made, after suitable shifts
of the world-lines, of two offending terms in the Lagrangian at the
order $G^4$. We shall therefore resort to two ambiguity parameters to
account for the different possible prescriptions regarding the IR
regularization at the 4PN order. Note that the presence of more than
one ambiguity parameter in our harmonic-coordinates approach has been
suggested in Ref.~\cite{DJS16}. As it turns out, the ambiguity
parameter that was introduced in Paper~I is equivalent to one linear
combination of them. In Ref.~\cite{BBBFM16c}, we shall specifically
employ the powerful dimensional regularization to handle the IR
divergences and investigate whether one can determine, with such
method, some combination of those ambiguity parameters. We emphasize
that they appear in very few terms of the 4PN Lagrangian, which
otherwise contains hundreds of difficult terms that have been
unambiguously determined in Paper~I.

Using this new Lagrangian modified by the two ambiguity parameters, we
use our new treatment of the non-locality in the dynamics to compute
the conserved integral of energy and the orbital precession of the
periastron at the 4PN order, in the limiting case of circular
orbits. We find that we can adjust the two ambiguity parameters in
such a way that the results are in agreement with the known GSF
calculations. Their values are uniquely fixed by this comparison,
which determines completely our Fokker Lagrangian. However, as said
above, we have to make an assumption regarding the structure of the
second ambiguity term. Work should thus continue in order to better
understand the origin of the ambiguity parameters. Comparing our 4PN
Lagrangian in harmonic coordinates with the 4PN Hamiltonian in
ADM-like coordinates as derived in
Refs.~\cite{JaraS12,JaraS13,JaraS15,DJS14,DJS16}, we find that there
exists a unique shift of the dynamical variables which connects the
two dynamics, thus our harmonic-coordinates Lagrangian is in fact
equivalent to the ADM Hamiltonian.

The plan of this paper is as follows. In Sec.~\ref{sec:ambiguity}, we
introduce two (and only two) ambiguity parameters to account for some
incompleteness in the IR regularization of integrals entering the
gravitational part of the Fokker action. In Sec.~\ref{sec:conserved},
we investigate the problem of defining conserved integrals of energy
and angular momentum from a Hamiltonian that is non-local in time. In
particular, we find that some constant (DC type) terms must be added
to the naive expectations for the energy and angular momentum. In
Sec.~\ref{sec:conserved}, we compute the tail contribution to the
energy at the 4PN order in the case of circular orbits. An alternative
derivation, which makes use of Delaunay variables, is also presented
(in Sec.~\ref{sec:delaunay}). In Sec.~\ref{sec:periK}, we compute the
periastron advance at the 4PN order in the circular orbit limit,
mostly focusing on the delicate tail contribution therein. The paper
ends with a short conclusion in Sec.~\ref{sec:conclusions}, followed
by several Appendices: A recapitulation of the complete 4PN
Lagrangians in Sec.~\ref{app:L}, some useful material about the
Fourier decomposition of the Newtonian quadrupole moment in
Sec.~\ref{app:fourier}, and complements on Sommerfeld's method of
contour integrals in Sec.~\ref{app:sommerfeld}.

\section{Ambiguity parameters in the Fokker Lagrangian}
\label{sec:ambiguity}

It was shown in Paper~I that infra-red (IR) divergences, due to the
behaviour of integrals at their bound at spatial infinity, start
appearing at the 4PN order in the Fokker action of point particles. As
it turned out, those IR divergences are associated with the presence
of tail effects~\cite{BD88,B93}.  In Paper~I, we found that the two
arbitrary scales associated with the tails and the IR cut-off scale
(denoted as $s_0$ and $r_0$, respectively, in Paper~I) combine, to
give a single ``ambiguity'' parameter $\alpha=\ln(r_0/s_0)$ which
could not be determined within the Fokker Lagrangian method. This
parameter was then fixed from a gravitational self-force (GSF)
calculation of the invariant energy for circular orbits.  Similar
results were obtained by means of the Hamiltonian formalism in
Ref.~\cite{DJS14}. The treatment of the IR divergences in Paper~I was
based on the use of Hadamard partie finie integrals. On the other
hand, the ultra-violet (UV) divergences associated with point
particles were handled by resorting to dimensional regularization.

For the present purpose, we first need to restore the arbitrariness of
the constant parameter $\alpha$ (which was fixed to $\alpha=811/672$
in Paper~I).  More precisely, starting from the 4PN
harmonic-coordinates Lagrangian of Paper~I, which is given by
Eqs.~(5.1)--(5.6) there, and notably including the non-local tail
term~(5.4), we reinstate $\alpha$ as an undetermined ambiguity
parameter by setting
\begin{equation}\label{L'}
L' = L^\text{Paper~I} + \frac{2G^2 m}{5 c^8}\left(\alpha -
\frac{811}{672}\right)\left(I^{(3)}_{ij}\right)^2\,,
\end{equation}
where $m=m_1+m_2$ is the total mass whereas $I^{(3)}_{ij}$ denotes the
third time derivative of the quadrupole moment, as given by Eq.~(5.12)
in Paper~I.  Next, we shall argue, based on preliminary
investigations~\cite{BBBFM16c}, that the problem of the
regularization procedure invoked to cure the IR divergences of the
Fokker action at the 4PN order is quite subtle. A choice was made in
Paper~I to regularize IR-divergent integrals by means of a specific
procedure based on analytic continuation in some complex parameter
$B$. Such a procedure (finite part when $B\to 0$) resulted in a
particular prescription for the gravitational part of the Fokker
Lagrangian, as given by Eq.~(2.20) in Paper~I. One can actually show
that this procedure is equivalent to the well known Hadamard Partie
Finie regularization~\cite{hadamard,schwartz} when applied to the
bound at spatial infinity.

However, preliminary calculations~\cite{BBBFM16c} suggest that
using the dimensional regularization instead of the Hadamard
regularization for the bound at infinity does change the content of
the Fokker Lagrangian (\textit{i.e.}, the associated gauge-invariant
quantities are different). Building on this observation, we shall
conjecture here that using different IR prescriptions entails a
modification of the Lagrangian by two types of terms, always having
the same structure. Such a behaviour is characteristic of the
appearance of ambiguities in the regularization process. This will be
acknowledged in the present paper by adding to the
Lagrangian~\eqref{L'} the ambiguous terms by hands, in agreement with
Ref.~\cite{DJS16} who pointed out that there might be a second
ambiguity parameter in our Fokker Lagrangian in harmonic
coordinates. In a future paper~\cite{BBBFM16c}, we shall
investigate whether specifically using the powerful dimensional
regularization should cure the IR divergences of the Fokker action in
a consistent way at the 4PN order, in addition to already dealing with
the UV divergences. In a first step, we add for convenience three
ambiguity parameters to~\eqref{L'}, $\beta_1$, $\beta_2$ and
$\beta_3$, which yields
\begin{equation}\label{L''}
L'' = L' + \frac{G^4 m \,m_1^2m_2^2}{c^8r_{12}^4}\left(\beta_1
(n_{12}v_{12})^2 + \beta_2 v_{12}^2 + \beta_3\frac{G
  m}{r_{12}}\right)\,.
\end{equation}
The positions and velocities of the particles are denoted by
$\bm{y}_A$ and $\bm{v}_A=\ud \bm{y}_A/\ud t$ ($A=1,2$);
$r_{12}=\vert\bm{y}_1-\bm{y}_2\vert$ is the relative separation (in
harmonic coordinates), $\bm{n}_{12}=(\bm{y}_1-\bm{y}_2)/r_{12}$ is the
corresponding unit vector, and $\bm{v}_{12}=\bm{v}_1-\bm{v}_2$ the
relative velocity. We use parenthesis to denote ordinary scalar
products, hence $(n_{12}v_{12})=\bm{n}_{12}\cdot\bm{v}_{12}$ and
$v_{12}^2=\bm{v}_{12}\cdot\bm{v}_{12}$. In a second step, inserting
into Eq.~\eqref{L'} the expression of the quadrupole moment $I_{ij}$
(to Newtonian order), we get
\begin{equation}\label{L'''}
L''' = L^\text{Paper~I} + \frac{G^4 m
  \,m_1^2m_2^2}{c^8r_{12}^4}\left(\gamma_1 (n_{12}v_{12})^2 + \gamma_2
v_{12}^2 + \gamma_3\frac{G m}{r_{12}}\right)\,,
\end{equation}
with $\gamma_1=\beta_1-\frac{176}{15}\alpha+\frac{8921}{630}$,
$\gamma_2=\beta_2+\frac{64}{5}\alpha-\frac{1622}{105}$ and
$\gamma_3=\beta_3$.  When evaluating the time derivatives of $I_{ij}$,
we actually replace $I_{ij}^{(3)}$ by its order reduced expression
(later denoted $\hat{I}_{ij}^{(3)}$), so that, in fact, $L'''$ differs
from $L''$ by a gauge transformation. There are still three ambiguity
parameters at this stage, but among the terms they generate one
combination is pure gauge. Thus, without loss of generality, we can
consider the following Lagrangian:
\begin{equation}\label{Lnew}
L = L^\text{Paper~I} + \frac{G^4 m
  \,m_1^2m_2^2}{c^8r_{12}^4}\Bigl(\delta_1 (n_{12}v_{12})^2 + \delta_2
v_{12}^2 \Bigr)\,,
\end{equation}
differing from $L'''$ by a further gauge transformation and containing
two (and only two) ambiguity parameters, related to the previous ones
by $\delta_1=\gamma_1-4\gamma_3$ and
$\delta_2=\gamma_2+\gamma_3$.\footnote{For completeness, let us
  mention that the gauge transformation of the Lagrangian (modulo an
  irrelevant total time derivative) has the following
  ``zero-on-shell'' form (with $a_A^i$ being the accelerations):
$$L = L''' - \sum_A m_A \left(a_A^i - (\partial_i
  U)_A\right)\xi_A^i\,,$$
where $U=\sum_B \frac{G m_B}{\vert\bm{x}-\bm{y}_B\vert}$. We have
$\xi_1^i = \gamma_3 \frac{G^4 m \,m_1 m_2^2}{c^8r_{12}^3}n_{12}^i$
(and $1 \leftrightarrow 2$) for the case at hand.} From the
Lagrangian~\eqref{Lnew}, one may construct, performing the various
local and non-local shifts of the particles world-lines, the
associated Hamiltonian in ADM type coordinates as in Paper~I. Since
those shifts represent small 2PN quantities at least, the extra terms
in~\eqref{Lnew}, of 4PN order, will be unchanged in the process.  The
resulting Hamiltonian reads
\begin{equation}\label{Hnew}
H = H^\text{Paper~I} - \frac{G^4 m
  \,m_1^2m_2^2}{c^8r_{12}^4}\Bigl(\delta_1 (n_{12}v_{12})^2 + \delta_2
v_{12}^2 \Bigr)\,,
\end{equation}
where the Hamiltonian of Paper~I, defined in Sec.~V C
of~\cite{BBBFM16a}, contains in particular the non-local tail term
given by Eq.~(5.17) of~\cite{BBBFM16a}. With an abuse of notation, we
employ the same lower-case letters as there to denote the ADM like
conjugate variables $\bm{y}_A$ and $\bm{p}_A$, with the obvious
shorthand notation $\bm{v}_A=\bm{p}_A/m_A$ applicable for the small
4PN extra terms in Eq.~\eqref{Hnew}. However, it is important to
beware that the variables of the formulations~\eqref{Lnew}
and~\eqref{Hnew} actually differ by appropriate shifts. With this
caveat in mind, we will often play indifferently with the Lagrangian
or Hamiltonian formalisms.

Now, in the present paper, we shall show that the two ambiguity
parameters $\delta_1$ and $\delta_2$ can be uniquely fixed by making
our dynamics compatible with existing GSF computations of the
conserved energy and periastron advance for circular orbits in the
small mass-ratio limit $\nu=(m_1m_2)/m^2\to 0$. Anticipating the
results of our computations, we shall get
\begin{equation}\label{ambparam}
\delta_1 = -\frac{2179}{315}\,,\qquad\delta_2 = \frac{192}{35}\,.
\end{equation}
Those values will be obtained (in Secs.~\ref{sec:Ecirc}
and~\ref{sec:periK}) after taking carefully into account the non-local
character of the 4PN tail term in the Lagrangian or Hamiltonian. With
the latter correcting terms and the specific values~\eqref{ambparam}
of the ambiguity parameters, our harmonic-coordinate 4PN dynamics
agrees with the 4PN Hamiltonian dynamics of
Refs.~\cite{JaraS12,JaraS13,JaraS15,DJS14,DJS16}. Indeed, building on
the work presented in Paper~I, we find that there exists a unique
(non-local-in-time) shift of the trajectories connecting the
harmonic-coordinate variables to the conjugate canonical ADM
Hamiltonian variables, so that our Hamiltonian~\eqref{Hnew} is
equivalent to the ADM Hamiltonian fully displayed in Appendix~A
of~\cite{DJS14}. Finally, we recapitulate in Appendix~\ref{app:L} our
results for the complete 4PN Lagrangian both in harmonic and ADM-like
coordinates.

\section{Conserved integrals for a non-local Hamiltonian}
\label{sec:conserved}

In this section, we investigate the notions of conserved energy and
angular momentum in the case of a non-local (in time) dynamics. For
convenience, we adopt the Hamiltonian formalism with the
Hamiltonian~\eqref{Hnew}. The latter is equivalent, after performing
some shifts of the variables, to the Lagrangian of
Sec.~\ref{sec:ambiguity}. We refer to Refs.~\cite{Llosa,Ferialdi} for
general discussions on non-local-in-time Hamiltonians.

In the Hamiltonian approach, the two-body system is described by the
canonical conjugate variables $\bm{y}_A$ and $\bm{p}_A$, with
$A=1,2$. (Again, for simplicity sake, we name these variables using
the same lower-case letters as in Sec.~\ref{sec:ambiguity}.) Those
canonical variables, in the center-of-mass frame, reduce to the
relative position of the particles, \textit{i.e.},
$\bm{x}=\bm{y}_1-\bm{y}_2$, and the linear momentum
$\bm{p}=\bm{p}_1=-\bm{p}_2$. We often denote $\bm{x}\equiv
(x^i)_{i=1,2,3}$ and $\bm{p}\equiv (p_i)_{i=1,2,3}$. We also pose, as
usual, $\bm{x}=r\bm{n}$ with $r=\vert\bm{x}\vert$ and $\bm{n}^2=1$,
whereas $p_r=\bm{n}\cdot\bm{p}$ represents the momentum conjugate to
$r$ in polar coordinates.

We shall consider the generic situation of interest for us where the
center-of-mass Hamiltonian is made of a local ``instantaneous'' piece
and a non-local-in-time ``tail'' part:
\begin{equation}\label{Hdecomp}
H[\bm{x}, \bm{p}] = H_0(\bm{x}, \bm{p}) + H^\text{tail}[\bm{x},
  \bm{p}]\,.
\end{equation}
The instantaneous piece $H_0$ is an ordinary local function of the
canonical variables $\bm{x}(t)$ and $\bm{p}(t)$, while the tail piece
is a \textit{functional} of the same variables, depending on
$\bm{x}(t+\tau)$ and $\bm{p}(t+\tau)$ for any
$\tau\in\mathbb{R}$. This dependence is indicated using squared
brackets. Furthermore, the functional is actually time symmetric. In
the specific case of the 4PN dynamics of compact binaries, the
instantaneous piece contains many PN contributions up to the 4PN
order. On the other hand, the non-local tail contribution arises at
the 4PN order and reads~\cite{FStail,DJS14,BBBFM16a,GLPR16}
\begin{equation}\label{tailterm}
H^\text{tail} = - \frac{G^2M}{5c^8}
\,\hat{I}_{ij}^{(3)}(t)\!\mathop{\text{Pf}}_{2r(t)/c}
\int_{-\infty}^{+\infty} \frac{\ud \tau}{\vert\tau\vert}
\,\hat{I}_{ij}^{(3)}(t+\tau)\,.
\end{equation}
This non-local tail piece ensures that the (conservative part of the)
4PN tail effect, known in the metric and equations of motion of
general matter systems, is recovered in the Hamiltonian framework
(see~\cite{BD88,B93}). The integral~\eqref{tailterm} involves the
symmetric kernel function $\mu(\tau)=1/\vert\tau\vert$ and possesses a
singular bound at $\tau=0$, which is handled with the Hadamard
``Partie finie'' (Pf). The partie finie depends on an arbitrary scale
(see, \textit{e.g.}, Ref.~\cite{BFreg}), chosen here to be the
separation distance between the two particles at the current time $t$,
\textit{i.e.}, $r(t)=\vert\bm{x}(t)\vert$. An alternative, more
explicit form of the tail integral is
\begin{equation}\label{tailtermlog}
H^\text{tail} = - \frac{G^2M}{5c^8} \,\hat{I}_{ij}^{(3)}(t)
\int_0^{+\infty} \ud s\ln\left(\frac{c \, s}{2
  r(t)}\right)\left[\hat{I}_{ij}^{(4)}(t-s) -
  \hat{I}_{ij}^{(4)}(t+s)\right]\,.
\end{equation}

In Eqs.~\eqref{tailterm}--\eqref{tailtermlog}, $M$ denotes the ADM
mass of the binary (which reduces to $m=m_1+m_2$ in the lowest
approximation) and $\hat{I}_{ij}^{(n)}(t)$ stands for the $n$-th
time-derivative of the Newtonian quadrupole moment of the system in
which the accelerations have been order-reduced by means of the
Newtonian equations of motion.  Following~\cite{BBBFM16a}, we indicate
the application of such a procedure of order reduction by adding a hat
on the concerned quantity. Thus, for instance, $\hat{I}_{ij}^{(3)}$
and $\hat{I}_{ij}^{(4)}$ are ordinary functions of the center-of-mass
canonical variables $x^i = r n^i$ and $p^i$, given by
\begin{subequations}\label{Iij34}
\begin{align}
\hat{I}_{ij}^{(3)} &= \frac{2G m}{r^2}\Bigl( 3 p_r n^{\langle
  i}n^{j\rangle} - 4 n^{\langle
  i}p^{j\rangle}\Bigr)\,,\label{Iij3}\\ \hat{I}_{ij}^{(4)} &=
\frac{2G}{r^3\nu}\biggl( \Bigl[3 \bm{p}^2 - 15 p_r^2+\frac{G
    m^3\nu^2}{r}\Bigr] n^{\langle i}n^{j\rangle} + 18 p_r n^{\langle
  i}p^{j\rangle} - 4 p^{\langle i}p^{j\rangle}\biggr)\,,\label{Iij4}
\end{align}\end{subequations}
with $\nu=m_1m_2/m^2$, the angular brackets denoting the symmetric-trace-free
(STF) projection. Note that, strictly speaking, since the accelerations are
order reduced in the two forms~\eqref{tailterm} and~\eqref{tailtermlog}, the
equivalence between~\eqref{tailterm} and~\eqref{tailtermlog} is true only
modulo a shift of the phase variables. Such shifts are generally ignored here.

The basic starting point is of course the dynamical action written in
Hamiltonian form,
\begin{equation}\label{action}
S = \int_{-\infty}^{+\infty} \ud t\left( p_i\frac{\ud x^i}{\ud t} - H
[\bm{x}, \bm{p}] \right)\,.
\end{equation}
Requiring that the action be stationary around the solution, we obtain
Hamilton's equations, in a form appropriate even for a non-local
Hamiltonian.  These will involve \textit{functional derivatives} with
respect to the canonical variables, hence\footnote{The functional
  derivatives should be more accurately denoted $\delta H/\delta
  p_i(t)$ and $\delta H/\delta x^i(t)$.  For any non-local functional
  $F[f]$ of a function $f(t)$, the functional derivative $\delta
  F/\delta f(t)$ is defined, after suitable integrations by parts
  (ignoring all integrated contributions at $t=\pm\infty$), by
\begin{equation}
\delta\int \ud t \,F(t) = \int \ud t \,\delta f(t)\,\frac{\delta
  F}{\delta f(t)}\nonumber\,.
\end{equation}}
\begin{equation}\label{HamiltonEOM}
\frac{\ud x^i}{\ud t} = \frac{\delta H}{\delta p_i}\,,\qquad\frac{\ud
  p_i}{\ud t} = - \frac{\delta H}{\delta x^i}\,.
\end{equation}
For future reference, we note that the functional derivatives of the tail part
of the Hamiltonian involve ordinary partial derivatives of the third
derivative of the quadrupole moment~\eqref{Iij3} with respect to the canonical
variables. More precisely, they read
\begin{subequations}\label{taileqs}
\begin{align}
\frac{\delta H^\text{tail}}{\delta x^i} &= -
\frac{2G^2M}{5c^8}\biggl[\frac{\partial \hat{I}_{jk}^{(3)}}{\partial
    x^i}\mathop{\text{Pf}}_{2r/c} \int_{-\infty}^{+\infty} \frac{\ud
    \tau}{\vert\tau\vert} \,\hat{I}_{jk}^{(3)}(t+\tau) -
  \frac{n^i}{r}(\hat{I}_{jk}^{(3)})^2\biggr]\,,\label{taileqsa}\\ \frac{\delta
  H^\text{tail}}{\delta p_i} &= - \frac{2G^2M}{5c^8}\frac{\partial
  \hat{I}_{jk}^{(3)}}{\partial p_i}\mathop{\text{Pf}}_{2r/c}
\int_{-\infty}^{+\infty} \frac{\ud \tau}{\vert\tau\vert}
\,\hat{I}_{jk}^{(3)}(t+\tau)\,.\label{taileqsb}
\end{align}
\end{subequations}
Notice the second term in Eq.~\eqref{taileqsa}, which comes from the
differentiation of the Hadamard partie finie scale chosen here to be $r(t)$.

\subsection{Integral of energy}
\label{sec:integralE}

We start by computing the time derivative of the Hamiltonian ``on shell'',
\textit{i.e.}, when the Hamiltonian equations of motion~\eqref{HamiltonEOM}
are satisfied by $x^i$ and $p_i$. Contrary to what happens in the usual local
case, the non-local Hamiltonian is not conserved on shell (even neglecting the
radiation reaction damping effects) because of the functional derivatives
present in the Hamiltonian equations~\eqref{HamiltonEOM}. Using
Eqs.~\eqref{taileqs}, we find instead the ``non-conservation'' law
\begin{equation}\label{dHdt}
\frac{\ud H}{\ud t} = \frac{G^2M}{5c^8}\left(\hat{I}_{ij}^{(4)}(t)
\mathop{\text{Pf}}_{2r/c} \int_{-\infty}^{+\infty} \frac{\ud
  \tau}{\vert\tau\vert}\,\hat{I}_{ij}^{(3)}(t+\tau) -
\hat{I}_{ij}^{(3)}(t)\mathop{\text{Pf}}_{2r/c}
\int_{-\infty}^{+\infty} \frac{\ud
  \tau}{\vert\tau\vert}\,\hat{I}_{ij}^{(4)}(t+\tau)\right)\,.
\end{equation}
It is worth mentioning that the Hadamard partie finie scale $r$
cancels out from the right-hand side of~\eqref{dHdt}. Moreover, even
though the non-local Hamiltonian on shell is not conserved, it is
actually conserved in an integrated sense, \textit{i.e.},
$\int_{-\infty}^{+\infty} \ud t\,\frac{\ud H}{\ud t} = 0$ (see
Refs.~\cite{Llosa,Ferialdi}). Furthermore one can easily check, using
for instance Eq.~(5.25) in Ref.~\cite{BBBFM16a}, that the right-hand
side of~\eqref{dHdt} is zero in the case of circular orbits.

We shall now construct from the law~\eqref{dHdt} the conserved energy
$E$ associated with the non-local Hamiltonian. To this end, we perform
a Taylor expansion of the integrand of~\eqref{dHdt} when $\tau\to
0$. Since the kernel function $\mu(\tau)=1/\vert\tau\vert$ is even, we
may consider only the even powers of $\tau$. However, the expansion
remains formal, since each of the coefficients of the Taylor expansion
is a divergent integral at the bounds $\tau=\pm\infty$. To remedy
this, we perform all our manipulations with the modified kernel
function $\mu(\tau)=\ue^{-\epsilon\vert\tau\vert}/\vert\tau\vert$ for
some $\epsilon>0$, and we will let $\epsilon$ tend to zero at the end
of our calculation to get a finite result.\footnote{In
  Ref.~\cite{DJS16}, the kernel function
  $\mu(\tau)=\frac{1}{\sqrt{2\pi\sigma^2}}\ue^{-\frac{1}{2}(\frac{\tau}{\sigma})^2}$
  was adopted.} At this stage, we can write
\begin{equation}\label{dHdtTaylor}
\frac{\ud H}{\ud t} = \frac{G^2M}{5c^8}
\sum_{n=1}^{+\infty}\left[\hat{I}_{ij}^{(4)}(t)\hat{I}_{ij}^{(2n+3)}(t) -
  \hat{I}_{ij}^{(3)}(t)\hat{I}_{ij}^{(2n+4)}(t)\right]\int_{-\infty}^{+\infty}
\frac{\ud \tau\,\ue^{-\epsilon\vert\tau\vert}}{\vert\tau\vert}
\frac{\tau^{2n}}{(2n)!}\,.
\end{equation}
We no longer need the Hadamard partie finie, because the integrals are
convergent at the bound $\tau=0$ for $n\geqslant 1$. Under the
form~\eqref{dHdtTaylor}, it is straightforward to recast the right-hand side
as a total time derivative. Indeed, one readily checks that
\begin{equation}\label{formderive}
\hat{I}_{ij}^{(4)}\hat{I}_{ij}^{(2n+3)} - \hat{I}_{ij}^{(3)}\hat{I}_{ij}^{(2n+4)} =
\frac{\ud}{\ud t}\left[-\hat{I}_{ij}^{(3)}\hat{I}_{ij}^{(2n+3)} +
  2\sum_{s=0}^{n-2}(-)^s \hat{I}_{ij}^{(s+4)}\hat{I}_{ij}^{(2n-s+2)} -
  (-)^n\bigl(\hat{I}_{ij}^{(n+3)}\bigr)^2\right]\,.
\end{equation}
This relation is valid for any $n\geqslant 1$, with the second term involving
the sum to be simply ignored when $n=1$. This formula proves that the
``non-conservation'' law~\eqref{dHdt} is in fact equivalent to a conservation
law \textit{stricto sensu}, namely $\ud E/\ud t = 0$. The conserved energy
$E$, though, differs from the Hamiltonian. It is defined instead by
$E=H+\delta H$, with
\begin{equation}\label{Eformal}
\delta H = \frac{G^2M}{5c^8}
\sum_{n=1}^{+\infty}\biggl[\hat{I}_{ij}^{(3)}\hat{I}_{ij}^{(2n+3)} -
  2\sum_{s=0}^{n-2}(-)^s \hat{I}_{ij}^{(s+4)}\hat{I}_{ij}^{(2n-s+2)} + (-)^n
  \bigl(\hat{I}_{ij}^{(n+3)}\bigr)^2\biggr]\!\int_{-\infty}^{+\infty}
\frac{\ud \tau\,\ue^{-\epsilon\vert\tau\vert}}{\vert\tau\vert}
\frac{\tau^{2n}}{(2n)!}\,.
\end{equation}
The quantity $E=H+\delta H$ actually represents the Noetherian conserved
energy associated with the Hamiltonian containing the non-local
term~\eqref{tailterm}. Obviously, the contribution of the tail term to that
energy is just $E^\text{tail}=H^\text{tail}+\delta H$, where $H^\text{tail}$
is given by Eq.~\eqref{tailterm}.

However, the expression~\eqref{Eformal} is not yet what we need since it is in
the form of an infinite Taylor expansion, while we would like to get a
non-perturbative expression for the conserved energy. The most convenient
approach to resum the Taylor series is to use a Fourier decomposition of the
binary dynamics at the Newtonian approximation (which is sufficient for our
purpose). The Newtonian quadrupole moment is a periodic function of time
(there is no orbital precession at this order), which may be decomposed as the
discrete Fourier series
\begin{equation}\label{Fourier}
I_{ij}(t) =
\sum_{p=-\infty}^{+\infty}\,\mathop{{\mathcal{I}}}_{p}{}_{\!\!ij}\,\ue^{\ui
  \,p\,\ell} \quad\text{with}\quad \mathop{{\mathcal{I}}}_{p}{}_{\!\!ij} =
\int_0^{2\pi}\frac{\ud\ell}{2\pi}\,I_{ij}\,\ue^{-\ui\, p\,\ell}\,,
\end{equation}
where $\ell=\omega(t-t_0)$ is the mean anomaly of the binary motion, with
$\omega=2\pi/P$ being the orbital frequency (or mean motion) corresponding to
the orbital period $P$, and $t_0$ the instant of passage at periastron. The
discrete Fourier coefficients ${}_{p}{\mathcal{I}}_{ij}$ are functions of the
orbit's eccentricity $e$ (to Newtonian order) and satisfy
${}_{p}{\mathcal{I}}_{ij}={}_{-p}{\overline{\mathcal{I}}}_{ij}$, with the
overbar denoting the complex conjugation. Their explicit expressions for
generic elliptic orbits in terms of combinations of Bessel functions depending
on the eccentricity are given in the Appendix A of~\cite{ABIQ08tail}. We
provide some alternative expressions in the Appendix~\ref{app:fourier} below.

Plugging~\eqref{Fourier} into~\eqref{Eformal}, we obtain a double Fourier
series indexed by integers $p$ and $q$, from which we separate out the
constant (DC) part corresponding to modes $p+q=0$ from the oscillating (AC)
part corresponding to $p+q\not= 0$. After some manipulations, we are able to
nicely resum the Taylor expansions when $\tau\to 0$ in both DC and AC parts to
simple trigonometric functions. We obtain
\begin{align}\label{Efourier}
\delta H =& - \frac{G^2M \omega^6}{5c^8}
\biggl[\sum_{p=-\infty}^{+\infty}\,\vert
  \mathop{{\mathcal{I}}}_{p}{}_{\!\!ij}\vert^2 p^6
  \int_{-\infty}^{+\infty} \frac{\ud
    \tau\,\ue^{-\epsilon\vert\tau\vert}}{\vert\tau\vert}
  \,(p\omega\tau)\sin(p\omega\tau) \\ & \qquad +\frac{1}{2}\sum_{p+q
    \not= 0}\,\mathop{{\mathcal{I}}}_{p}{}_{\!\!ij}
  \mathop{{\mathcal{I}}}_{q}{}_{\!\!ij} \,\frac{p^3q^3(p-q)}{p+q}
  \,\ue^{\ui(p+q)\ell} \int_{-\infty}^{+\infty} \frac{\ud
    \tau\,\ue^{-\epsilon\vert\tau\vert}}{\vert\tau\vert}
  \,\bigl[\cos(p\omega\tau) - \cos(q\omega\tau)\bigr] \biggr]
\,.\nonumber
\end{align}
The remaining integrals are convergent at the bound $\tau=0$, as well
as at infinity $\tau=\pm\infty$ thanks to our exponential cut-off
factor $\ue^{-\epsilon\vert\tau\vert}$. Moreover, they can be
evaluated with standard formulas,\footnote{Namely,
\begin{align}
\int_{-\infty}^{+\infty} \frac{\ud
  \tau\,\ue^{-\epsilon\vert\tau\vert}}{\vert\tau\vert}
\,(p\omega\tau)\sin(p\omega\tau) &=
2\,,\nonumber\\\int_{-\infty}^{+\infty} \frac{\ud \tau
  \,\ue^{-\epsilon\vert\tau\vert}}{\vert\tau\vert}\,\bigl[\cos(p\omega\tau)
  - \cos(q\omega\tau)\bigr] &= - 2
\ln\left|\frac{p}{q}\right|\,,\nonumber
\end{align}
where we have set $\epsilon=0$ after the integration.\label{formulas}}
yielding
\begin{equation}\label{Efinal}
\delta H = - \frac{2G^2M \omega^6}{5c^8}
\biggl[\sum_{p}\,\vert\mathop{{\mathcal{I}}}_{p}{}_{\!\!ij}\vert^2 p^6
  - \frac{1}{2} \sum_{p+q \not=
    0}\,\mathop{{\mathcal{I}}}_{p}{}_{\!\!ij}\mathop{{\mathcal{I}}}_{q}{}_{\!\!ij}
  \,\frac{p^3q^3(p-q)}{p+q}
  \ln\left|\frac{p}{q}\right|\,\ue^{\ui(p+q)\ell}\biggr]\,.
\end{equation}
This is our final result for $E=H+\delta H$, valid for general binary orbits.
There are no other contributions to this conserved integral of energy
associated to the non-local Hamiltonian $H=H_0+H^\text{tail}$ at the 4PN
order. Notice that the DC contribution [first term in~\eqref{Efinal}] admits a
closed analytic form in physical space. We see that it is indeed directly
related to the leading order total averaged energy flux
$\mathcal{F}_\text{GW}$ emitted in gravitational waves (with $\langle\rangle$
referring to the time averaging):
\begin{equation}\label{EDCpart}
\delta H^\text{DC} = -
\frac{2G^2M}{5c^8}\,\langle\bigl(\hat{I}_{ij}^{(3)}\bigr)^2\rangle = -
\frac{2G M}{c^3}\,\mathcal{F}_\text{GW}\,.
\end{equation}

The second term in~\eqref{Efinal}, with oscillating modes $p+q \not=
0$, can straightforwardly be computed. Its expression may be checked
by a direct integration of the Fourier decomposition of the right-hand
side of the ``non-conservation'' law~\eqref{dHdt}. However, such a
direct integration misses an integration constant which cannot
\textit{a priori} be guessed by this method. Its correct value is
provided by the constant DC contribution, \textit{i.e.}, the first
term in~\eqref{Efinal} (corresponding to Fourier modes with
$p+q=0$). Finally, the important point for us is that, while for
circular orbits the AC term vanishes, the DC term does not. In the
circular orbit limit, the non-zero modes reduce to the quadrupolar
ones ($p=\pm 2$), and to the modes $p=0$ which will not contribute
here.\footnote{\label{ft8} We have (with $m$ being the total mass and
  $\nu$ the symmetric mass ratio)
\begin{align}
&\mathop{{\mathcal{I}}}_{2}{}_{\!\!xx} =
  \mathop{{\mathcal{I}}}_{-2}{}_{\!\!xx} = \frac{1}{4}m\,\nu r^2\,,
  &\mathop{{\mathcal{I}}}_{2}{}_{\!\!yy} =
  \mathop{{\mathcal{I}}}_{-2}{}_{\!\!yy} = -\frac{1}{4}m\,\nu
  r^2\,,\nonumber \\[-0.2cm] &\mathop{{\mathcal{I}}}_{2}{}_{\!\!xy} =
  -\mathop{{\mathcal{I}}}_{-2}{}_{\!\!xy} = -\frac{\ui}{4}m\,\nu r^2\,,
  &\mathop{{\mathcal{I}}}_{0}{}_{\!\!xx} =
  \mathop{{\mathcal{I}}}_{0}{}_{\!\!yy} =
  -\frac{1}{2}\mathop{{\mathcal{I}}}_{0}{}_{\!\!zz} = \frac{1}{6}m\,\nu
  r^2\,.\nonumber
\end{align}
} Eq.~\eqref{Efinal} becomes then (with $M=m$)
\begin{equation}\label{Efinalcirc}
\delta H = - \frac{64}{5} m c^2\,\nu^2 x^5\,.
\end{equation}
As usual we have posed $x=(G m \omega/c^3)^{2/3}$. The extra
term~\eqref{Efinalcirc} was not considered in Paper~I. As we shall see,
including it gladly resolves our disagreement with the derivation of the
conserved energy for circular orbits in the Hamiltonian
formalism~\cite{DJS14}.

\subsection{Integral of angular momentum}
\label{sec:integraleJ}

We keep considering the problem of defining conserved quantities from a
non-local-in-time Hamiltonian, but focusing on the angular momentum. We assume
that the instantaneous piece $H_0$ of the center-of-mass
Hamiltonian~\eqref{Hdecomp} is rotationally invariant, \textit{i.e.}, depends
on the canonical variables only through the rotational invariants
$r=\vert\bm{x}\vert$, $p_r=\bm{n}\cdot\bm{p}$ and $\bm{p}^2$. This is surely
the case for our 4PN Hamiltonian describing the dynamics of compact binaries
without spins. Let us therefore study the impact of the non-local tail piece
of $H$ on the conservation law of the angular momentum. First of all, it is
straightforward to derive the law of variation of the ``orbital'' angular
momentum $\bm{L}=\bm{x}\times\bm{p}$ (or $L^i=\varepsilon_{ijk} x^j p^k$)
which would be conserved in the absence of the tail term:
\begin{equation}\label{dLdt0}
\frac{\ud L^i}{\ud t} = - \varepsilon_{ijk}\left(x^j\,\frac{\delta
  H^\text{tail}}{\delta x^k} + p^j\,\frac{\delta H^\text{tail}}{\delta
  p^k} \right)\,.
\end{equation}
Using the functional derivatives~\eqref{taileqs} of the tail part of the
Hamiltonian and the explicit expression of the third derivative of the
quadrupole moment~\eqref{Iij3}, we further obtain
\begin{equation}\label{dLdt}
\frac{\ud L^i}{\ud t} =
\frac{4G^2M}{5c^8}\,\varepsilon_{ijk}\,\hat{I}_{jl}^{(3)}
\mathop{\text{Pf}}_{2r/c}\int_{-\infty}^{+\infty} \frac{\ud
  \tau}{\vert\tau\vert} \,\hat{I}_{kl}^{(3)}(t+\tau)\,,
\end{equation}
noticing again the cancellation of the Hadamard partie finie scale on the
right-hand side. Next, proceeding as we did for the energy, we Taylor expand
the integrand when $\tau\to 0$, after adding as previously the regulator
$\ue^{-\epsilon\vert\tau\vert}$ to avoid divergence problems:
\begin{equation}\label{dLdtexpand}
\frac{\ud L^i}{\ud t} = \frac{4G^2M}{5c^8}\,\varepsilon_{ijk}
\sum_{n=1}^{+\infty}\hat{I}_{jl}^{(3)}\hat{I}_{kl}^{(2n+3)}\int_{-\infty}^{+\infty}
\frac{\ud \tau\,\ue^{-\epsilon\vert\tau\vert}}{\vert\tau\vert}
\frac{\tau^{2n}}{(2n)!}\,.
\end{equation}
It is no longer necessary to keep the Hadamard partie finie on the remaining
integral. Now, the appropriate analogue of the formula~\eqref{formderive} is
\begin{equation}\label{formderive2}
\varepsilon_{ijk}\hat{I}_{jl}^{(3)}\hat{I}_{kl}^{(2n+3)} = \frac{\ud}{\ud
  t}\left[\sum_{s=0}^{n-1}(-)^s \varepsilon_{ijk}
  \hat{I}_{jl}^{(s+3)}\hat{I}_{kl}^{(2n-s+2)}\right]\,.
\end{equation}
Hence we obtain the full-fledged conservation law for the integral of angular
momentum: $\ud J^i/\ud t = 0$. When the Hamiltonian is non-local, $J^i = L^i +
\delta L^i$ contains, in addition to the naive guess $L^i=\varepsilon_{ijk}
x^j p^k$, the extra contribution
\begin{equation}\label{Jformal}
\delta L^i = - \frac{4G^2M}{5c^8}\sum_{n=1}^{+\infty}
\biggl[\sum_{s=0}^{n-1}(-)^s
  \varepsilon_{ijk}\hat{I}_{jl}^{(s+3)}\hat{I}_{kl}^{(2n-s+2)}\biggr]\int_{-\infty}^{+\infty}
\frac{\ud \tau\,\ue^{-\epsilon\vert\tau\vert}}{\vert\tau\vert}
\frac{\tau^{2n}}{(2n)!}\,.
\end{equation}
The Taylor expansion when $\tau\to 0$ can be summed up at the prize of
introducing Fourier series. We end up with the following result, composed of a
constant DC term and a non constant oscillation AC term:
\begin{equation}\label{Jfinal}
\delta L^i = \frac{4 G^2M \omega^5}{5c^8} \biggl[\sum_{p}
  \ui\varepsilon_{ijk} \mathop{{\mathcal{I}}}_{p}{}_{\!\!jl}
  \mathop{{\mathcal{I}}}_{-p}{}_{\!\!kl}\,p^5 - \sum_{p+q \not=
    0}\ui\varepsilon_{ijk}\mathop{{\mathcal{I}}}_{p}{}_{\!\!jl}
  \mathop{{\mathcal{I}}}_{q}{}_{\!\!kl} \,\frac{p^3q^3}{p+q}
  \ln\left|\frac{p}{q}\right|\,\ue^{\ui(p+q)\ell}\biggr]\,.
\end{equation}
For the DC term, a relation similar to Eq.~\eqref{EDCpart} for the energy
holds, namely
\begin{equation}\label{LDCpart}
(\delta L^i)^\text{DC} = \frac{4G^2 M}{5c^8}\,\langle
  \varepsilon_{ijk}\hat{I}_{jl}^{(3)}\hat{I}_{kl}^{(2)}\rangle = -
  \frac{2G M}{c^3}\,\mathcal{G}^i_\text{GW}\,,
\end{equation}
where $\mathcal{G}^i_\text{GW}$ denotes the total (averaged) flux of angular
momentum at leading PN order. Finally, for circular orbits, the latter
expression reduces to the DC contribution
\begin{equation}\label{Jfinalcirc}
\delta \bm{L} = - \frac{64}{5} \frac{G m^2}{c}\nu^2 x^{7/2} \bm{\ell}
= \delta p_\varphi \bm{\ell} \,,
\end{equation}
with $\bm{\ell}=\bm{L}/\vert\bm{L}\vert$, the symbol $\delta p_\varphi$
denoting the corresponding contribution in the restricted problem (see
Sec.~\ref{sec:consE}), such that $h=p_\varphi+\delta p_\varphi$ is
conserved.

\section{Conserved energy for circular orbits}
\label{sec:Ecirc}

In the present section, we shall compute the energy $E$ for
circular orbits, extending the derivation of Paper~I by including the
extra contribution $\delta H^\text{DC}$ investigated in the previous
section. In the next section~\ref{sec:periK}, we will also compute the
advance of the periastron in the circular orbit limit, but we shall find
that, in this case, the extra contributions in the energy and the
angular momentum do not contribute.

When the orbits are exactly circular, \textit{i.e.}, when the
space-time admits a helical Killing vector, both the energy and
periastron advance can be determined in the small mass-ratio limit
using perturbative gravitational self-force
methods~\cite{Detweilerorleans,Barackorleans}. One first obtains the
redshift observable~\cite{Det08}, which is the invariant associated
with the helical Killing symmetry. From this quantity, one deduces the
energy using the first law of binary point-particle
mechanics~\cite{LBW12}. To get the periastron advance, one needs both
an averaged version of the redshift observable~\cite{BarackS11} and a
generalization of the first law valid for eccentric
orbits~\cite{L15,BL16}. We shall thus limit ourselves to the case of
circular orbits in order to make meaningful comparisons between our
4PN results and the self-force results.

\subsection{Derivation using reduced canonical variables}
\label{sec:consE}

We consider the restricted problem of particle motion in a fixed orbital
plane, which is appropriate for the 4PN dynamics of particles without spin.
Let $\varphi$ be the phase angle or true anomaly of the orbit, defined from
the orbit's periastron. For the restricted problem, the relative separation is
given by $\bm{x}=r\bm{n}$ with the unit vector
$\bm{n}=(\cos\varphi,\sin\varphi)$ in polar coordinates. Introducing the
second unit vector $\bm{\lambda}=(-\sin\varphi,\cos\varphi)$, orthogonal to
$\bm{n}$ in the orbital plane, we also decompose $\bm{p}=p_r
\bm{n}+p_\varphi\bm{\lambda}/r$, where $p_\varphi$ is the conserved angular
momentum in the case of a rotationally invariant Hamiltonian. Then,
$\bm{\ell}=\bm{n}\times\bm{\lambda}$ is the unit vector normal to the orbital
plane and we have $\bm{L}=\bm{x}\times\bm{p}=p_\varphi\bm{\ell}$. Another
useful notation is to pose
$\bm{m}=\frac{1}{\sqrt{2}}(\bm{n}+\ui\,\bm{\lambda})$ (see
Appendix~\ref{app:fourier}).

Performing the change of canonical variables $(\bm{x},
\bm{p})\rightarrow (r,\varphi,p_r,p_\varphi)$, Hamilton's
equations, derived from the action $S=\int [p_r\ud
  r+p_\varphi\ud\varphi - H \ud t]$, read
\begin{subequations}\label{Heqsrest}
\begin{align}
\frac{\ud r}{\ud t} &= \frac{\delta H}{\delta p_r}\,,\qquad\frac{\ud
  p_r}{\ud t} = - \frac{\delta H}{\delta
  r}\,,\label{Heqsresta}\\ \frac{\ud \varphi}{\ud t} &= \frac{\delta
  H}{\delta p_\varphi}\,,\qquad\frac{\ud p_\varphi}{\ud
  t} = - \frac{\delta H}{\delta \varphi}\,,\label{Heqsrestb}
\end{align}
\end{subequations}
where the derivatives are still to be considered in a functional
sense.
%\footnote{One can verify that $(r, \varphi, p_r, p_\varphi)$ are
%  canonical variables even when the Hamiltonian is non local.}
 The Hamiltonian, as in Eq.~\eqref{Hdecomp}, is again the sum of the
 instantaneous piece $H_0(r,p_r,p_\varphi)$ and of the non-local tail
 term~\eqref{tailterm}. While the instantaneous piece $H_0$ is
 independent of the phase angle $\varphi$, the tail piece does depend
 on it, so that the conjugate momentum $p_\varphi$ is no longer
 conserved for the non-local dynamics [see
   Eqs.~\eqref{dLdt0}--\eqref{dLdt}].

In this section, we shall mainly study the contribution of the tail term
in the circular energy, as the computations due
to the instantaneous part are standard. We simply choose for the
instantaneous term the Newtonian Hamiltonian and add to it the tail
term~\eqref{tailterm}, setting
\begin{equation}\label{Htot}
H = \frac{1}{2m\nu}\left(p_r^2+\frac{p_\varphi^2}{r^2}\right) - \frac{G
  m^2\nu}{r} + H^\text{tail}\,.
\end{equation}
We then write the tail term as $H^\text{tail}=-(G^2M)/(5c^8)\,
\hat{I}_{ij}^{(3)}\mathcal{T}_{ij}^{(3)}$ with the following special
notation for the tail factor (where the Hadamard Pf is still defined
with the scale $r$)
\begin{equation}\label{tailfactor}
\mathcal{T}_{ij}^{(a)} = \mathop{\text{Pf}}_{2r/c}
\int_{-\infty}^{+\infty} \frac{\ud \tau}{\vert\tau\vert}
\,\hat{I}_{ij}^{(a)}(t+\tau)\,.
\end{equation}
This tail factor may be computed explicitly by means of the Fourier
series~\eqref{Fourier}, which leads to (with $\gamma_\text{E}$ the Euler
constant)
\begin{equation}\label{tailfactorFourier}
\mathcal{T}_{ij}^{(a)} = -2 \sum_{p=-\infty}^{+\infty} (\ui
p\,\omega)^a\,\mathop{{\mathcal{I}}}_{p}{}_{\!\!ij} \,\ue^{\ui p
  \ell}\biggl(\ln\left(\frac{2\vert p\vert\omega r}{c}\right) +
\gamma_\text{E} \biggr)\,.
\end{equation}

In the following, we shall systematically project out the tail factor
onto the moving vector basis $(\bm{n},\bm{\lambda})$, thus defining
$\mathcal{T}_{nn}^{(a)} = n^in^j\mathcal{T}_{ij}^{(a)}$,
$\mathcal{T}_{n\lambda}^{(a)} = n^i\lambda^j\mathcal{T}_{ij}^{(a)}$
and $\mathcal{T}_{\lambda\lambda}^{(a)} =
\lambda^i\lambda^j\mathcal{T}_{ij}^{(a)}$. For instance, thanks
to~\eqref{Iij3}, we may put the tail piece of the Hamiltonian in the
form (with $M=m$ in this approximation)
\begin{equation}\label{Htailexpl}
H^\text{tail} = \frac{2G^3m^2}{5c^8r^2} \left[ p_r
  \mathcal{T}_{nn}^{(3)} +
  \frac{4p_\varphi}{r}\mathcal{T}_{n\lambda}^{(3)}\right]\,.
\end{equation}
The Hamilton's equations~\eqref{Heqsrest}, in which we consistently
use the functional derivatives of the tail part of the Hamiltonian,
can be written more explicitly as
\begin{subequations}\label{Heqsdetail}
\begin{align}
\frac{\ud r}{\ud t} &= \frac{p_r}{m\nu} + \frac{4G^3m^2}{5c^8r^2}
\mathcal{T}_{nn}^{(3)} \,,\label{Heqsdetaila}\\\frac{\ud p_r}{\ud t}
&= \frac{p_\varphi^2}{m\nu r^3} - \frac{G m^2\nu}{r^2} +
\frac{8G^3m^2}{5c^8r^3}
\left[p_r\mathcal{T}_{nn}^{(3)}+\frac{6p_\varphi}{r}
  \mathcal{T}_{n\lambda}^{(3)}-\frac{Gm}{r^2} \left(\frac{2}{3}p_r^2 +
  8\frac{p_\varphi^2}{r^2}\right)\right]\,,\label{Heqsdetailb}\\ \frac{\ud
  \varphi}{\ud t} &= \frac{p_\varphi}{m\nu r^2} +
\frac{16G^3m^2}{5c^8r^3} \mathcal{T}_{n\lambda}^{(3)}
\,,\label{Heqsdetailc}\\\frac{\ud p_\varphi}{\ud t} &=
-\frac{8G^3m^2}{5c^8r^2} \left[p_r\mathcal{T}_{n\lambda}^{(3)} +
  \frac{2p_\varphi}{r}\left(\mathcal{T}_{\lambda\lambda}^{(3)} -
  \mathcal{T}_{nn}^{(3)}\right)\right]\,.\label{Heqsdetaild}
\end{align}
\end{subequations}

Let us now derive the tail contribution to the conserved energy for circular
orbits (extending the derivation from Sec.~VD in Paper~I). Imposing $p_r=0$ in
Eqs.~\eqref{Heqsdetail}, we first recover the usual Newtonian solution
$p_\varphi=p_\varphi^0+\mathcal{O}(c^{-8})$, $\varphi=\omega_0
t+\mathcal{O}(c^{-8})$, in which $p_\varphi^0=m\nu\sqrt{G m r_0}$ and
$\omega_0=\sqrt{G m/r_0^3}$ are the constant Newtonian angular momentum and
orbital frequency, respectively, with $r_0$ being the radius of the circular
orbit. Proceeding iteratively, we inject the Newtonian solution into the tail
terms of Eqs.~\eqref{Heqsdetail}. These can then be reduced for circular
orbits with the help of the formulas [see Eq.~(5.25) in Paper~I]:
\begin{subequations}\label{evaluatetail}
\begin{align}
\left(\mathcal{T}_{n\lambda}^{(3)}\right){\Big|}_{(p_\varphi^0,\omega_0
  t)} &=
8\frac{G^{3/2}m^{5/2}\nu}{r^{5/2}}\biggl[\ln\left(\frac{4\omega_0
    r_0}{c}\right) +
  \gamma_\text{E}\biggr]\,,\\ 
\left(\mathcal{T}_{nn}^{(3)}\right){\Big|}_{(p_\varphi^0,\omega_0
  t)} &=
\left(\mathcal{T}_{\lambda\lambda}^{(3)}\right){\Big|}_{(p_\varphi^0,\omega_0
  t)} = 0 \,.
\end{align}
\end{subequations}
In turn, the system of equations~\eqref{Heqsdetail} including tails
admits the solutions $p_\varphi=p_\varphi^0+\Delta p_\varphi$ and
$\varphi=(\omega_0+\Delta\omega)t$, where $\Delta p_\varphi$ and
$\Delta\omega$ are the tail contributions to the angular momentum and
orbital frequency, respectively, as a function of the radius $r_0$. Their
expressions are:
\begin{subequations}\label{Deltatail}
\begin{align}
\Delta p_\varphi(r_0) &=
\frac{G^{9/2}m^{11/2}\nu^2}{c^8r_0^{7/2}}\biggl[-\frac{192}{5}
  \biggl(\ln\left(\frac{4\omega_0 r_0}{c}\right) +
  \gamma_\text{E}\biggr)+\frac{32}{5}\biggr]\,,\label{Deltataila}\\ 
\Delta\omega(r_0) &=
\frac{G^{9/2}m^{9/2}\nu}{c^8r_0^{11/2}}\biggl[-\frac{64}{5}
  \biggl(\ln\left(\frac{4\omega_0 r_0}{c}\right) +
  \gamma_\text{E}\biggr)+\frac{32}{5}\biggr]\,.\label{Deltatailb}
\end{align}
\end{subequations}
This solves two of Hamilton's equations, while the two others simply
say that $r_0$ and $p_\varphi$ are constant for circular orbits.

We are now in a position to calculate the conserved energy associated with the
non-local Hamiltonian for circular orbits. As shown in
Sec.~\ref{sec:conserved}, one must take into account extra contributions in
the definition of the energy and angular momentum with respect to the naive
expectations, \textit{i.e.}, $H$ and $\bm{L}=\bm{x}\times\bm{p}$,
respectively. In fact, we have seen that $E=H+\delta H$ and
$\bm{J}=\bm{L}+\delta\bm{L}$, with $\delta H$ and $\delta\bm{L}$ given
by~\eqref{Efinal} and~\eqref{Jfinal}, or by the explicit
formulas~\eqref{Efinalcirc} and~\eqref{Jfinalcirc}, respectively, in the
circular orbit limit. For the restricted problem, since
$\bm{L}=p_\varphi\bm{\ell}$, we shall decompose the true angular momentum
invariant $h$ as $h=p_\varphi+\delta p_\varphi$, where $\delta p_\varphi$ is
given by~\eqref{Jfinalcirc} for circular orbits. Beware that $\Delta
p_\varphi$ denotes the tail contribution in $p_\varphi$, whereas $\delta
p_\varphi$ serves to connect $p_\varphi$ to the angular momentum invariant
$h$, as appropriate for our non-local Hamiltonian.

Thus, the conserved energy for circular orbit is the sum of the total
Hamiltonian $H$, made of the instantaneous part as well as the tail
term~\eqref{Htailexpl}, which is to be evaluated thanks to
Eqs.~\eqref{evaluatetail}, and of the crucial term $\delta H$ given
by~\eqref{Efinalcirc}. After using~\eqref{Deltataila}, we find that the tail
contribution of the circular energy, regarded as a function of $r_0$, is
\begin{equation}\label{Er}
\Delta E(r_0) =
\frac{G^{5}m^{6}\nu^2}{c^8r_0^{5}}\biggl[-\frac{128}{5}
  \biggl(\ln\left(\frac{4\omega_0 r_0}{c}\right) +
  \gamma_\text{E}\biggr)-\frac{32}{5}\biggr] \,.
\end{equation}
Finally, the invariant energy must be expressed in terms of the
orbital frequency through the usual parameter $x=(G m
\omega/c^3)^{2/3}$. We obtain 
\begin{equation}\label{Ex} 
\Delta E(x) = - \frac{224}{15} m c^2 \nu^2 x^5\biggl[\ln\left(16
  x\right) + 2\gamma_\text{E} + \frac{2}{7}\biggr] \,.
\end{equation}
The above value for $\Delta E(x)$ includes notably the tail
contribution due to the replacement of $r_0$ as a function of $x$ into
the Newtonian energy $E_\text{N} = -\frac{1}{2}Gm^2\nu/r_0$ by means
of Eq.~\eqref{Deltatailb}. The result~\eqref{Ex} corrects Eq.~(5.30)
in Paper~I and we are now in agreement with the Hamiltonian formalism
of Refs.~\cite{DJS14,DJS16}, regarding the treatment of nonlocalities
in the circular energy. It is the extra term $\delta H$ displayed in
Eq.~\eqref{Efinalcirc} that permits reconciling our work with the
Hamiltonian results.

The complete expression of the circular energy through the 4PN order
is the sum of the tail part~\eqref{Ex} and of all the terms coming
from the instantaneous part of the 4PN dynamics. The latter terms may
be computed either in Lagrangian form, using the harmonic coordinates
Lagrangian of Paper~I augmented by the two ambiguity parameters
$\delta_1$ and $\delta_2$ as advocated in Sec.~\ref{sec:ambiguity}, or
in Hamiltonian form. After reducing to the frame of the center of mass
and specializing to circular orbits, we get the full 4PN result, which
however still depends on one of the ambiguity parameters, namely
$\delta_2$. As usual, the ambiguous parameter is expected to be a pure
rational fraction entering the coefficient that is linear in the
symmetric mass ratio $\nu$, since the terms involving logarithms or
irrationals such as $\pi$ and $\gamma_\text{E}$ are uniquely
determined. To find this fraction, we compare our expression with the
result of GSF calculations, valid in the small mass-ratio
limit~\cite{BDLW10b,BiniD13}. This uniquely fixes
$\delta_2=\frac{192}{35}$. In the end, our complete invariant 4PN
circular energy as a function of the orbital frequency reads
\begin{align}\label{Ecirc4PN}
	E &= -\frac{m\nu c^2 x}{2} \biggl\{ 1 + \left( - \frac{3}{4} -
        \frac{\nu}{12} \right) x + \left( - \frac{27}{8} +
        \frac{19}{8} \nu - \frac{\nu^2}{24} \right) x^2 \nonumber
        \\ &\qquad\quad + \left( - \frac{675}{64} + \biggl[
          \frac{34445}{576} - \frac{205}{96} \pi^2 \biggr] \nu -
        \frac{155}{96} \nu^2 - \frac{35}{5184} \nu^3 \right) x^3
        \nonumber \\ &\qquad\quad + \left( - \frac{3969}{128} +
        \left[-\frac{123671}{5760}+\frac{9037}{1536}\pi^2 +
          \frac{896}{15}\gamma_\text{E}+ \frac{448}{15} \ln(16
          x)\right]\nu\right.\nonumber\\ & \qquad\qquad\quad \left.+
        \left[-\frac{498449}{3456}+\frac{3157}{576}\pi^2\right]\nu^2
        +\frac{301}{1728}\nu^3 + \frac{77}{31104}\nu^4\right) x^4
        \biggr\} \,.
\end{align}

\subsection{Derivation using Delaunay-Poincar\'e variables}
\label{sec:delaunay}
 
In this section, we present a derivation based on a simple non-local
model that has been proposed in Refs.~\cite{DJS16}. Moreover, we aim
at comparing and contrasting the non-local dynamics to some other
simpler models also discussed in~\cite{DJS16}. Starting from the
original 4PN non-local action for compact binaries, an alternative
local action has been derived in~\cite{DJS16} by applying non-local
shifts of the particle world-lines and removing ``double-zero''
contributions. This local action has then been used in applications
such as building the 4PN effective-one-body (EOB)
Hamiltonian~\cite{DJS15eob}. In this section, we shall check that, in
such a local dynamics, the conserved energy for circular orbits
reproduces that of the original non-local dynamics. Furthermore, we
shall confirm that the so-called ``Ostrogradski'' dynamics
investigated in~\cite{DJS16} is equivalent to this local dynamics. It
is thus also equivalent to the non-local dynamics, at least regarding
the evaluation of the conserved energy for circular orbits.  Since
Ref.~\cite{DJS16} insisted on the usefulness of a particular set of
variables, known as the elliptical Delaunay variables, we shall adopt
those variables here.

Starting from the restricted problem (the motion lying in a fixed
orbital plane), using the canonical variables $(\bm{x},\bm{p})$, we
define the velocity as $\bm{v}=\bm{p}/\mu$ (where $\mu=m\nu$) and
parametrize the motion by means of the usual elliptic elements,
\textit{i.e.}, the semi-major axis $a$, the eccentricity $e$, the
argument of the periastron $g$ (often denoted $\omega$ in celestial
mechanics), and the instant of passage at periastron $t_0$. The
Delaunay variables are then given by $(\ell, g, \mu \mathcal{L}, \mu
\mathcal{G})$, where $\ell=\omega(t-t_0)$ is the mean anomaly,
$\omega=(G m/a^3)^{1/2}$ is the mean motion (generally denoted $n$ in
celestial mechanics), $\mathcal{L}=(G m a)^{1/2}$, and $\mathcal{G}=[G
  m a(1-e^2)]^{1/2}$ is the angular momentum \textit{per} unit
mass. The important point is that the variables $(\ell, g, \mu
\mathcal{L}, \mu \mathcal{G})$ are canonical, with $\ell$ and $g$
representing the generalized positions, $\mu\mathcal{L}$ and
$\mu\mathcal{G}$ being the associated generalized momenta. In fact, we
shall use a variant of these variables, known as the Poincar\'e
canonical variables, $(\lambda, h, \Lambda, \mathcal{H})$, where the
generalized positions are $\lambda=\ell+g$ and $h=-g$, while the
associated canonical momenta are $\Lambda=\mu\mathcal{L}$ and
$\mathcal{H}=\mu(\mathcal{L}-\mathcal{G})$. Their main advantage is
that $\lambda$ represents directly the phase variable, measured from a
fixed direction in the orbital plane (thus taking into account the
effect of orbital precession), so that $\omega=\dot{\lambda}$ is
nothing but the orbital frequency.

For our discussion we keep only the Newtonian term and the 4PN tail
term given by~\eqref{tailterm}. Furthermore, one advantage of the
Delaunay-Poincar\'e variables (as emphasized in~\cite{DJS16}) is that
we can treat the tail term at the level of the Hamiltonian as an
expansion when the eccentricity tends to zero, where
\begin{equation}\label{e}
e(\Lambda, \mathcal{H}) =
\frac{\sqrt{\mathcal{H}\left(2\Lambda-\mathcal{H}\right)}}{\Lambda}\,.
\end{equation}
As we are ultimately interested in the energy for circular orbits, we
shall neglect the eccentricity in the Hamiltonian. On the other hand,
notice that the semi-major axis is only a function of $\Lambda$
through
\begin{equation}\label{a}
a(\Lambda) = \frac{\Lambda^2}{G m\mu^2}\,.
\end{equation}
Thus, $\Lambda$ will be directly related to the radius of the circular
orbit. Working out the non-local Hamiltonian~\eqref{Htot} in the limit
of $e\to 0$ and using the Delaunay-Poincar\'e variables we obtain
\begin{equation}\label{Hnonlocal0}
H = -\frac{G^2 m^2 \mu^3}{2\Lambda^2} -
\frac{32G^{10}m^{9}\mu^{12}}{5c^8}\left(\mathop{\text{Pf}}_{2\sqrt{\Lambda_0}}
\int_{-\infty}^{+\infty}\frac{\ud
  \tau}{\vert\tau\vert}\,\frac{\cos[2(\lambda-\lambda')]}
    {\Lambda^5\Lambda'^5} - 4
    \frac{\ln(\Lambda/\Lambda_0)}{\Lambda^{10}}\right) +
    \mathcal{O}\left(e\right)\,.
\end{equation}
We denote $\lambda=\lambda(t)$ and $\lambda'=\lambda(t+\tau)$ together
with $\Lambda=\Lambda(t)$, $\Lambda'=\Lambda(t+\tau)$. When $e\to 0$
the Hamiltonian depends functionally only on the phase variable
$\lambda$ and its conjugate momentum $\Lambda$. We have a Hadamard
partie finie in front of the non-local integral, and the last term
in~\eqref{Hnonlocal0} accounts for the fact that the partie finie
depends on a scale which is the separation at time $t$, \textit{i.e.},
$r=a(1-e \cos u)$ in elliptical representation, which yields
$r\propto\sqrt{\Lambda}$ in the limit $e\to 0$. The constant
$\Lambda_0$ introduced in~\eqref{Hnonlocal0} is irrelevant and
actually cancels out from the two terms.

We introduce a small parameter $\varepsilon\propto c^{-8}$ in front of
the tail term and work at linear order in $\varepsilon$. Posing
$Gm=\nu=c=\Lambda_0=1$ for simplicity, and ignoring the remainder term
in the eccentricity and the Hadamard partie finie (always implicitly
understood in what follows), we have the very simple model
\begin{equation}\label{Hnonlocal}
H = -\frac{1}{2\Lambda^2} +
\varepsilon\left(\int_{-\infty}^{+\infty}\frac{\ud
  \tau}{\vert\tau\vert}\,
\frac{\cos[2(\lambda-\lambda')]}{\Lambda^5\Lambda'^5} - 4
\frac{\ln\Lambda}{\Lambda^{10}}\right) \,.
\end{equation}
The dynamics follows from varying the action $S = \int[ \Lambda \ud
  \lambda - H \ud t]$. The non-locality results in functional
derivatives in Hamilton's equations, hence
\begin{subequations}\label{eqsHnonlocal}
\begin{align}
\dot{\lambda} &= \frac{\delta H }{\delta \Lambda} =
\frac{1}{\Lambda^3} +
\varepsilon\left(-10\,\int_{-\infty}^{+\infty}\frac{\ud
  \tau}{\vert\tau\vert}\,
\frac{\cos[2(\lambda-\lambda')]}{\Lambda^6\Lambda'^5} +
\frac{40\ln\Lambda-4}
     {\Lambda^{11}}\right)\,,\label{eqsHnonlocala}\\\dot{\Lambda} &= -
     \frac{\delta H }{\delta \lambda} =
     4\varepsilon\,\int_{-\infty}^{+\infty}\frac{\ud
       \tau}{\vert\tau\vert} \,\frac{\sin[2(\lambda-\lambda')]}
     {\Lambda^5\Lambda'^5}\,.\label{eqsHnonlocalb}
\end{align}\end{subequations}

We shall now solve those equations for circular orbits. Noticing that the
orbital frequency $\omega=\dot{\lambda}$ of the circular motion is
equal to $1/\Lambda^3$ to zero-th order in $\varepsilon$, and that
$\Lambda$ is constant at that order, we can solve the equations thanks
to the results [see Eq.~(5.25) in Paper~I]
\begin{subequations}\label{intsincos}
\begin{align}
\int_{-\infty}^{+\infty}\frac{\ud
  \tau}{\vert\tau\vert}\,\sin[2(\lambda-\lambda')] &=
0\,,\\ \int_{-\infty}^{+\infty}\frac{\ud
  \tau}{\vert\tau\vert}\,\cos[2(\lambda-\lambda')] &= -2\left[
  \ln\left(\frac{4}{\Lambda^3}\right) + \gamma_\text{E} \right]\,.
\end{align}\end{subequations}
The equations~\eqref{eqsHnonlocal} are thus equivalent to saying that
$\Lambda$ is now constant to linear order in $\varepsilon$, and that
the following relation between $\Lambda$ and the orbital frequency
holds,
\begin{equation}\label{OmLambda}
\omega = \frac{1}{\Lambda^3} + \frac{20\varepsilon}{\Lambda^{11}}
\left[\ln\left(\frac{4}{\Lambda}\right) + \gamma_\text{E} -
  \frac{1}{5}\right]\,.
\end{equation}

Our next problem is to relate the conserved energy of the circular
orbit to the Hamiltonian~\eqref{Hnonlocal} computed on shell. This
problem has been solved in the general case in
Sec.~\ref{sec:integralE} and here we redo the reasoning within the
toy model~\eqref{Hnonlocal}. Differentiating~\eqref{Hnonlocal} with
the help of the equations of motion~\eqref{eqsHnonlocal} we get
\begin{equation}\label{dHdtDelaunay}
\frac{\ud H}{\ud t} = 2\varepsilon \int_{-\infty}^{+\infty}\frac{\ud
  \tau}{\vert\tau\vert}\bigl(\dot{\lambda}+\dot{\lambda}'\bigr)
\,\frac{\sin[2(\lambda-\lambda')]}{\Lambda^5\Lambda'^5} \,.
\end{equation}
As in Sec.~\ref{sec:integralE} we perform a formal Taylor expansion
when $\tau\to 0$ and are able to rewrite the right-hand side in the
form of a total derivative which is then transferred to the left-hand
side. This yields the looked-for energy in the form $E=H+\delta H$
where
\begin{align}\label{deltaHformal}
\delta H =& -\varepsilon \sum_{n=1}^{+\infty}\biggl[C\,C^{(2n)} -
  2\sum_{s=0}^{n-2}(-)^s C^{(s+1)}C^{(2n-s-1)} + (-)^n
  \bigl(C^{(n)}\bigr)^2\biggr]\!\int_{-\infty}^{+\infty} \frac{\ud
  \tau\,\ue^{-\epsilon\vert\tau\vert}}{\vert\tau\vert}
\frac{\tau^{2n}}{(2n)!}\nonumber\\ & + ~\bigl(\text{\textit{idem} with
  $C\longleftrightarrow S$}\bigr)\,.
\end{align}
We have posed $C = \cos(2\lambda)/\Lambda^5$ and $S =
\sin(2\lambda)/\Lambda^5$ and the second term represents exactly the
same expression but with $S$ in place of $C$. The Taylor expansion can
be re-summed up by going to Fourier space. We decompose $C$ and $S$ in
Fourier series [with the conventions~\eqref{Fourier}] and get
closed-form expressions involving trigonometric integrals that are
finally computed using Footnote~\ref{formulas}. Finally we have
\begin{align}\label{CSfinal}
\delta H &=
\varepsilon\biggl[2\sum_{p}\,\vert\mathop{\mathcal{C}}_{p}\vert^2 +
  \sum_{p+q \not=
    0}\,\mathop{\mathcal{C}}_{p}\mathop{{\mathcal{C}}}_{q}\,\frac{p-q}{p+q}
  \ln\left|\frac{p}{q}\right|\,\ue^{\ui(p+q)\ell}\biggr]\nonumber\\ &
+ ~\bigl(\text{\textit{idem} with
  $\mathop{{\mathcal{C}}}_{p}\longleftrightarrow
  \mathop{{\mathcal{S}}}_{p}$}\bigr)\,.
\end{align}
This is made of oscillatory terms having $p+q\not= 0$ and of a crucial
constant DC contribution with $p+q=0$. For circular orbits\footnote{In
  this case we just have
  ${}_2\mathcal{C}=\frac{1}{2\Lambda^5}={}_{-2}\mathcal{C}$ and
  ${}_2\mathcal{S}=\frac{1}{2\ui\Lambda^5}=-{}_{-2}\mathcal{S}$
  (others are zero).} we find that the DC term does give the extra
contribution $\delta H=2\varepsilon/\Lambda^{10}$ and we obtain, after
reducing Eq.~\eqref{Hnonlocal} to circular orbits
using~\eqref{intsincos},
\begin{equation}\label{ELambda}
E = - \frac{1}{2\Lambda^2} -
\frac{2\varepsilon}{\Lambda^{10}}\left[\ln\left(\frac{4}{\Lambda}\right)
  + \gamma_\text{E} - 1\right]\,.
\end{equation}
Combining~\eqref{OmLambda} and~\eqref{ELambda} we can express $E$ at
linear order in $\varepsilon$ as a function of the circular orbit
frequency, or, rather, the usual parameter $x=\omega^{2/3}$. The
result is
\begin{equation}\label{Enonlocal}
E = - \frac{x}{2} + \frac{7\varepsilon}{3} x^5
\left[\ln\left(16x\right) + 2\gamma_\text{E} + \frac{2}{7}\right]\,,
\end{equation}
which is consistent with Eq.~\eqref{Ex}, and with the result from the
ADM Hamiltonian approach obtained in Refs.~\cite{DJS14,DJS16}.

In fact, the latter works~\cite{DJS14,DJS16} proceeded in another
way. Instead of working with the original non-local
Hamiltonian~\eqref{Hnonlocal}, they first inserted into the non-local
tail integral the equations of motion ``off shell'', \textit{i.e.},
containing non-zero source terms $S_\lambda$ and $S_\Lambda$ defined
by the variation of the action with respect to the canonical
variables, $S_\lambda=\delta S/\delta\lambda$ and $S_\Lambda=\delta
S/\delta\Lambda$. In Eqs.~(4.7) of~\cite{DJS16} they obtained a
solution to those off-shell equations at linear order in the source
terms $S_\lambda$ and $S_\Lambda$, which they inserted back into the
non-local action, again keeping only the linear terms when
$S_\lambda,\,S_\Lambda\to 0$, arguing that the higher non-linear terms
are double-zero or multiple-zero contributions that do not affect the
dynamics. Then the extra terms linear in the source contributions
$S_\lambda$ and $S_\Lambda$ could be ``field-redefined away'' by some
shifts of the dynamical variables, say $\delta\lambda=\xi_\lambda$ and
$\delta\Lambda=\xi_\Lambda$, where the shifts are given as some
non-local integrals over $S_\lambda$ and $S_\Lambda$. Thus, according
to Ref.~\cite{DJS16}, performing these shifts justifies the naive
replacement of $\lambda'$ and $\Lambda'$ in the non-local
Hamiltonian~\eqref{Hnonlocal} by the solution of the Newtonian
equations of motion, which in this case is
$\lambda'=\lambda+\tau/\Lambda^3$ and $\Lambda'=\Lambda$. As a result
the non-local dynamics becomes local in the shifted variables, with the
Hamiltonian given by
\begin{equation}\label{Hlocal}
H = -\frac{1}{2\widetilde{\Lambda}^2} +
\varepsilon\left(\frac{1}{\widetilde{\Lambda}^{10}}
\,\int_{-\infty}^{+\infty}\frac{\ud \tau}{\vert\tau\vert}
\cos\!\left(\frac{2\tau}{\widetilde{\Lambda}^3}\right) - 4
\frac{\ln\widetilde{\Lambda}}{\widetilde{\Lambda}^{10}}\right) \,,
\end{equation}
where the shifted variables are
$\widetilde{\lambda}=\lambda+\xi_\lambda$ and
$\widetilde{\Lambda}=\Lambda+\xi_\Lambda$. This is the toy model of
Eq.~(4.12) in~\cite{DJS16}, except that here we keep the same last
term as in~\eqref{Hnonlocal} to account for our choice for the
Hadamard partie finie scale. Since~\eqref{Hlocal} depends only on
$\widetilde{\Lambda}$ and not on $\widetilde{\lambda}$ (the dependence
on $\widetilde{\lambda}$ would arise in eccentricity-dependent terms),
we have $\widetilde{\Lambda}=$ const, while the orbital frequency
comes from the conjugate Hamiltonian equation as
\begin{align}\label{OmegaLambdalocal}
\omega = \frac{\partial H}{\partial \widetilde{\Lambda}}
&= \frac{1}{\widetilde{\Lambda}^3} +
\varepsilon\left[-\frac{10}{\widetilde{\Lambda}^{11}}
  \,\int_{-\infty}^{+\infty}\frac{\ud
    \tau}{\vert\tau\vert}\cos\!\left(\frac{2\tau}
        {\widetilde{\Lambda}^3}\right)
        \right.\nonumber\\ &\left.\qquad\qquad~ +
        \frac{6}{\widetilde{\Lambda}^{14}}\,\int_{-\infty}^{+\infty}\frac{\ud
          \tau}{\vert\tau\vert}\,\tau\,\sin\!\left(\frac{2\tau}
             {\widetilde{\Lambda}^3}\right) +
             \frac{40\ln\widetilde{\Lambda}-4}
                  {\widetilde{\Lambda}^{11}}\right]\,.
\end{align}
Note that the second term in the square brackets is due to the
derivation of the ``frequency'' $1/\widetilde{\Lambda}^3$ in the
argument of the cosine inside the integral, and this contribution is
evaluated using the formulas in Footnote~\ref{formulas}. On the other
hand, in the local model there is no need to add an extra contribution
to the conserved energy $E$, since the energy is obviously given by
the Hamiltonian on shell. The extra term $\delta H$ in the non-local
model gives a contribution which exactly accounts for the presence of
the second term in Eq.~\eqref{OmegaLambdalocal}. Finally, taking into
account the relation between $\omega$ and $\widetilde{\Lambda}$
deduced from~\eqref{OmegaLambdalocal}, we obtain for the invariant
circular energy expressed in terms of $x=\omega^{2/3}$ the same result
as Eq.~\eqref{Enonlocal}. In particular this is a confirmation of the
procedure of ``localization'' of the non-local action by means of
non-local shifts, as advocated in Ref.~\cite{DJS16}.

The authors of~\cite{DJS16} also argued that the non-local dynamics is
equivalent to another dynamics, specified by a generalized Hamiltonian
\textit{\`a la} Ostrogradski obtained from the non-local
Hamiltonian~\eqref{Hnonlocal} by formally performing the Taylor
expansion $\lambda'=\lambda+\dot{\lambda}\tau+\mathcal{O}(\tau^2)$ and
keeping only the linear term, thus
\begin{equation}\label{Hostro}
H = -\frac{1}{2\Lambda^2} +
\varepsilon\left(\frac{1}{\Lambda^{10}}
\,\int_{-\infty}^{+\infty}\frac{\ud
  \tau}{\vert\tau\vert}\cos\bigl(2\dot{\lambda}\tau\bigr) -
4 \frac{\ln\Lambda}{\Lambda^{10}}\right) \,.
\end{equation}
This Hamiltonian is a generalized one, as it depends on
$\Lambda$ and $\dot{\lambda}$, but it is local
since the canonical variables are evaluated at time $t$. The
Hamiltonian equations read now
\begin{equation}\label{eqsHostro}
\dot{\lambda} = \frac{\partial H}{\partial
  \Lambda}\,,\qquad\dot{\Lambda} = \frac{\ud}{\ud
  t}\left(\frac{\partial H}{\partial
  \dot{\lambda}}\right)\,,
\end{equation}
while the associated conserved energy is given by
\begin{equation}\label{EostroH}
E = H -
\dot{\lambda}\frac{\partial H}{\partial
  \dot{\lambda}}\,.
\end{equation}
We readily find that the equations of motion imply that $\Lambda=$
const and that $\omega=\dot{\lambda}$ is related to $\Lambda$ by the
same relation~\eqref{OmLambda} as in the non-local model. Furthermore
the result is the same as for the non-local model because the second
term present in the conserved energy~\eqref{EostroH} plays the same
role as the extra term due to the DC part of Eq.~\eqref{CSfinal} in
the non-local model, and we end up again with the same result as in
Eq.~\eqref{Enonlocal}, thus confirming the arguments of
Ref.~\cite{DJS16}.

\section{Periastron advance for circular orbits}
\label{sec:periK}

We denote by $K$ the fractional angle of the periastron advance, such
that the precession of the periastron \textit{per} revolution is $\Phi
= 2\pi K$. As $K$ tends to one in the limit $c\to +\infty$ (no
precession at the Newtonian level) the relativistic precession is
entirely described by $k=K-1$. Because of the non-local tail
term~\eqref{tailterm}, it will be possible to compute $K$ only in the
form of an expansion series when the orbit's eccentricity $e\to 0$. In
fact we shall restrict our calculation to the limiting case of
circular orbits. In this section we shall focus our attention mostly on
the control of the tail contribution to the periastron. The
contributions due to the local part of the dynamics, will essentially
be dealt with using standard methods~\cite{DS88}, but some complements are
given in Appendix~\ref{app:sommerfeld}.

Even if we consider \textit{in fine} the limiting case of circular
orbits, we must be careful about taking into account high enough
corrections in the eccentricity, because of cancellations occurring
with powers of $e$ in denominators, yielding a finite result when
$e\to 0$. The precession equation we need to integrate in the
Hamiltonian formalism [see Eqs.~\eqref{Heqsrest}] is
\begin{equation}\label{dphidr}
\frac{\ud \varphi}{\ud r} = \frac{\delta H/\delta p_\varphi}{\delta
  H/\delta p_r}\,.
\end{equation}
In the more detailed model $H=H_0+H^\text{tail}$ where $H_0$ is the
Newtonian Hamiltonian, see~\eqref{Htot}, we insert
Eqs.~\eqref{Heqsdetail} into~\eqref{dphidr} and obtain, after
expanding to 4PN order,
\begin{equation}\label{dphidrinitial}
\frac{\ud \varphi}{\ud r} = \frac{p_\varphi}{r^2 p_r}\left[ 1 +
  \frac{4G^3m^3\nu}{5c^8r^2}\left(\frac{4r}{p_\varphi}
  \mathcal{T}_{n\lambda}^{(3)} -
  \frac{1}{p_r}\mathcal{T}_{nn}^{(3)}\right)\right]\,.
\end{equation}
Since we must consistently include high-order corrections in the
eccentricity, recall that $p_\varphi$ and the Hamiltonian $H$ are not
constant but oscillate at the 4PN order, and we must take into account
those variations to the precession
equation~\eqref{dphidrinitial}. With the notation~\eqref{tailfactor},
the equations obeyed by these quantities, which are equivalent
to~\eqref{dHdt} and~\eqref{dLdt}, read as
\begin{subequations}\label{Hpphi}
\begin{align}
\frac{\ud H}{\ud t} &= \frac{2G^2 m^2}{5c^8} \left[\frac{1}{m\nu r^3}
  \left(\left[2p_r^2+\frac{3h^2}{r^2} +
    \frac{m^3\nu^2}{r}\right]\mathcal{T}_{nn}^{(3)} + \frac{10h
    p_r}{r}\mathcal{T}_{n\lambda}^{(3)} -
  \frac{4h^2}{r^2}\mathcal{T}_{\lambda\lambda}^{(3)}\right)
  \right.\nonumber\\&\left.\qquad\qquad\quad + \frac{1}{r^2}
  \left(p_r\mathcal{T}_{nn}^{(4)} +
  \frac{4h}{r}\mathcal{T}_{n\lambda}^{(4)}\right)\right]\,,\\ \frac{\ud
  p_\varphi}{\ud t} &= -\frac{8G^3m^2}{5c^8r^2}
\left[p_r\mathcal{T}_{n\lambda}^{(3)}+\frac{2h}{r}
  \left(\mathcal{T}_{\lambda\lambda}^{(3)}-\mathcal{T}_{nn}^{(3)}\right)\right]\,.
\end{align}
\end{subequations}
In Sec.~\ref{sec:conserved} we related the solutions of these
equations to the conserved integrals of energy $E$ and angular
momentum $\bm{J}=h\bm{\ell}$, with results $E=H+\delta H$ and
$h=p_\varphi+\delta p_\varphi$ where $\delta H$ and $\delta p_\varphi$
have been provided in a general way by Eqs.~\eqref{Efinal}
and~\eqref{Jfinal}, with the obvious correspondence
$\bm{J}=\bm{L}+\delta\bm{L}=(p_\varphi+\delta
p_\varphi)\bm{\ell}$. Recall that $\delta H$ and $\delta p_\varphi$
contain both oscillating AC and constant DC terms. We have checked
that the DC terms in~\eqref{Efinal} and~\eqref{Jfinal} do not
contribute to the periastron advance. On the other hand the AC terms
play an important role and can be computed either from~\eqref{Efinal}
and~\eqref{Jfinal} or directly by integrating out Eqs.~\eqref{Hpphi},
using the Fourier transform of the tail
term~\eqref{tailfactorFourier}. Note that the Fourier series is
equivalent to an expansion when the eccentricity $e\to 0$ so it must
be pushed far enough to fully control the periastron even in the
circular orbit limit $e=0$.

Next we equate $H$ given by~\eqref{Htot} to $E-\delta H$, which
permits solving for the radial momentum $p_r$ consistently to the 4PN
order, with the result
\begin{equation}\label{pr}
p_r = f(r) \left[ 1 + \frac{h}{r^2 f^2(r)}\delta p_\varphi -
  \frac{m\nu}{f^2(r)}\left(H^\text{tail} + \delta H \right)\right]\,,
\end{equation}
in which $H^\text{tail}$ can also be replaced by the more explicit
expression~\eqref{Htailexpl}. Here we have introduced the function
$f(r)$ representing the Newtonian approximation for the radial
momentum in terms of the orbit's invariants $E$ and $h$, namely
\begin{equation}\label{fr}
f(r) = \sqrt{2m\nu E + \frac{2Gm^3\nu^2}{r} - \frac{h^2}{r^2}}\,.
\end{equation}
With such a notation we end up with a completely explicit expression for
the precession equation, depending also on the previously determined
$\delta H$ and $\delta p_\varphi$,
\begin{align}\label{dphidrinitial2}
\frac{\ud \varphi}{\ud r} = \frac{h}{r^2 f(r)}\Biggl\{ 1 &+ \frac{2
  G^3m^3\nu}{5c^8r^2}\left(-\frac{1}{f(r)}\mathcal{T}_{nn}^{(3)} +
\frac{4r}{h}\left[2+\frac{h^2}{r^2
    f^2(r)}\right]\mathcal{T}_{n\lambda}^{(3)}\right)\nonumber\\ & -
\biggl[1+\frac{h^2}{r^2 f^2(r)}\biggr]\frac{\delta p_\varphi}{h} +
\frac{m\nu}{f^2(r)}\delta H \Biggr\}\,.
\end{align}
In this expression the tail terms can be obtained using the Fourier
expansion~\eqref{tailfactorFourier}, together with usual formulas for
the Fourier decomposition of the Keplerian motion.\footnote{Among
  these let us report the well-known expansion of the eccentric
  anomaly $u$, related to the mean anomaly $\ell$ by Kepler's equation
  $\ell=u-e\sin{u}$:
\begin{align}
u &= \ell + 2\sum_{q=1}^{+\infty}\frac{1}{q}J_q(q e)\sin(q
\ell)\,,\nonumber\\[-0.2cm] \ue^{\ui k u} &=
-\frac{e}{2}\delta_{1,\vert k\vert} + \sum_{s\not= 0}\frac{k}{s}
J_{s-k}(s e) \,\ue^{\ui k \ell}\quad\text{(for any
  $k\in\mathbb{N}^*$)}\,.\nonumber
\end{align}
We provide in Appendix~\ref{app:fourier} the explicit expressions of
the Fourier coefficients of the Newtonian quadrupole moment in terms
of Bessel functions of the eccentricity.} Note that because the radial
function $f(r)$ is proportional to $e$ for a solution of the motion,
one must expand the tail terms (and $\delta H$, $\delta p_\varphi$ as
well) to sufficiently high order in $e$. Finally we integrate the
precession equation~\eqref{dphidrinitial2} between the turning points
$r_p$ and $r_a$ corresponding to periastron and apastron. All the
integrals can be computed either from Sommerfeld's method of complex
contour integrals or, equivalently, by using Hadamard's partie finie
method~\cite{DS88}. One of the integrals to be computed contains a
logarithm and we explain in Appendix~\ref{app:sommerfeld} how we
employ the method of contour integrals in this case. We obtain in the
limiting circular case $e\to 0$ the tail contribution to the
periastron advance as
\begin{equation}\label{Ktailcirc}
K^\text{tail} = \left(\frac{352}{5} - \frac{1256}{15}\ln x -
\frac{592}{15}\ln 2 - \frac{1458}{5}\ln 3 -
\frac{2512}{15}\gamma_\text{E}\right)\nu x^4\,.
\end{equation}

Next we evaluate the numerous instantaneous contributions up to 4PN
order. We did two independent calculations, one based on the
invariants associated with our harmonic-coordinates Lagrangian, as
given in Paper~I but corrected by Eq.~\eqref{G4} above, and one based
on the associated ADM Hamiltonian. The simplification of course, is
that with the instantaneous part of the dynamics the invariants are
defined in the usual way --- there is no need to worry about their
variations as in Eqs.~\eqref{Hpphi}, and the calculation can be done
for any eccentric orbits. Again we compute all integrals by means of
Sommerfeld's contour integrals~\cite{Sommerfeld} or alternatively by
Hadamard's partie finie integrals. As is well known, in harmonic
coordinates we meet integrals containing some logarithms. In that case
the Sommerfeld method is no longer straightforward and has to be
adapted as described in Appendix~\ref{app:sommerfeld}.

Considering the circular orbit limit and adding the tail
part~\eqref{Ktailcirc} we look for the values of the ambiguity
parameters $\delta_1$ and $\delta_2$ that are needed to reproduce the
known GSF contribution to the periastron advance when $\nu\to
0$~\cite{BDS10,Letal11,vdM16,D10sf,DJS16}. Like for the energy we find
that the logarithms and all irrationals (like $\pi^2$ and
$\gamma_\text{E}$) are already correct, and we get only one constraint
$\delta_1+14\delta_2=\frac{22013}{315}$. Since we already know that
$\delta_2=\frac{192}{35}$ from the circular limit of the energy (see
Sec.~\ref{sec:consE}), we obtain $\delta_1=-\frac{2179}{315}$. The
ambiguity parameters are therefore determined, as announced in
Eq.~\eqref{ambparam}. Thus the GSF limit has permitted us to uniquely
fix the values of the two ambiguity parameters and our 4PN dynamics is
now complete (for any mass ratio).

For the instantaneous part of the periastron advance and for general
orbits we find\footnote{Our computations make extensive use of the
software Mathematica together with the tensor package
\textit{xAct}~\cite{xtensor}.}
\begin{align}\label{Kinst}
K_0 &= 1 + \frac{3}{c^2\hat{h}^2} \nonumber\\&+
\frac{1}{c^4}\left[\frac{1}{\hat{h}^4}\left(\frac{105}{4} -
  \frac{15}{2}\nu\right)+\frac{\hat{E}}{\hat{h}^2}\left(\frac{15}{2} -
  3\nu\right)\right] \nonumber\\&+
\frac{1}{c^6}\left[\frac{1}{\hat{h}^6}\left(\frac{1155}{4}
  +\left(-\frac{625}{2}+\frac{615}{128}\pi^2\right)\nu+
  \frac{105}{8}\nu^2\right)\right.\nonumber\\&\left.\qquad +
  \frac{\hat{E}}{\hat{h}^4}\left(\frac{315}{2}
  +\left(-218+\frac{123}{64}\pi^2\right)\nu+ \frac{45}{2}\nu^2\right)
  + \frac{\hat{E}^2}{\hat{h}^2}\left(\frac{15}{4} - \frac{15}{4}\nu +
  3\nu^2\right)\right]
\nonumber\\&+\frac{1}{c^8}\left[\frac{1}{\hat{h}^8}
  \left(\frac{225225}{64} +\left(
  -\frac{1736399}{288}+\frac{2975735}{24576}\pi^2\right)\nu +\left(
  \frac{132475}{96}-\frac{7175}{256}\pi^2\right)\nu^2
  -\frac{315}{16}\nu^3\right) \right.\nonumber\\&\left.\qquad +
  \frac{\hat{E}}{\hat{h}^6} \left(\frac{45045}{16}
  +\left(-\frac{293413}{48}+\frac{257195}{2048}\pi^2\right)\nu
  +\left(\frac{35065}{16}-\frac{615}{16}\pi^2\right)\nu^2-
  \frac{525}{8}\nu^3\right)\right.\nonumber\\&\left.\qquad +
  \frac{\hat{E}^2}{\hat{h}^4}\left(\frac{4725}{16}
  +\left(-\frac{20323}{24}+\frac{35569}{2048}\pi^2\right)\nu +
  \left(\frac{4045}{8}-\frac{615}{128}\pi^2\right)\nu^2 -
  45\nu^3\right)\right.\nonumber\\&\left.\qquad +
  \frac{\hat{E}^3}{\hat{h}^2}\left(\frac{15}{4}\nu^2 -
  3\nu^3\right)\right]\,,
\end{align}
where we have conveniently rescaled the invariants as
$\hat{E}=\frac{E}{m\nu c^2}$ and $\hat{h}=\frac{h c}{G
  m^2\nu^2}$. Reducing~\eqref{Kinst} to circular orbits is quite
simple, as we need $\hat{E}$ and $\hat{h}$ in terms of $x=(\frac{G
  m\omega}{c^3})^{2/3}$ only to 3PN order (since the Newtonian term is
merely one), and we get
\begin{align}\label{Kinstcirc}
K_0 &= 1 + 3 x + \left(\frac{27}{2} - 7\nu\right) x^2 +
\left(\frac{135}{2}
+\left[-\frac{649}{4}+\frac{123}{32}\pi^2\right]\nu+ 7\nu^2\right) x^3
\nonumber\\& + \left(\frac{2835}{8}
+\left[-\frac{60257}{72}+\frac{48007}{3072}\pi^2\right]\nu+
\left[\frac{5861}{12}-\frac{451}{32}\pi^2\right]\nu^2 -
\frac{98}{27}\nu^3\right) x^4\,.
\end{align}
Finally, our complete prediction for the periastron advance in the
case of circular orbits is
\begin{align}\label{Ktotal}
K &= 1 + 3 x + \left(\frac{27}{2} - 7\nu\right) x^2 +
\left(\frac{135}{2}
+\left[-\frac{649}{4}+\frac{123}{32}\pi^2\right]\nu+ 7\nu^2\right) x^3
\nonumber\\& + \left(\frac{2835}{8}
+\left[-\frac{275941}{360}+\frac{48007}{3072}\pi^2 -
  \frac{1256}{15}\ln x - \frac{592}{15}\ln 2 - \frac{1458}{5}\ln 3 -
  \frac{2512}{15}\gamma_\text{E}\right]\nu
\right.\nonumber\\&\left.\qquad+
\left[\frac{5861}{12}-\frac{451}{32}\pi^2\right]\nu^2 -
\frac{98}{27}\nu^3\right) x^4\,,
\end{align}
which agrees with the result of~\cite{DJS15eob} obtained via the EOB
link derived in~\cite{D10sf}. The GSF contribution therein is
generally described by means of the function $\rho(x)$ such that
$K^{-2} = 1 - 6 x + \nu\rho(x) +
\mathcal{O}(\nu^2)$~\cite{D10sf,Letal11}, and we get
\begin{align}\label{rho}
\rho &= 14 x^2 + \left(\frac{397}{2}-\frac{123}{16}\pi^2\right) x^3
\nonumber\\& + \left(-\frac{215729}{180} + \frac{58265}{1536}\pi^2 +
\frac{1184}{15}\ln 2 + \frac{2916}{5}\ln 3 +
\frac{5024}{15}\gamma_\text{E} + \frac{2512}{15}\ln x\right) x^4\,.
\end{align}
The 4PN coefficient is of the type
$\rho_\text{4PN}=a_\text{4PN}+b_\text{4PN}\ln x$ and in particular,
the coefficient $a_\text{4PN}$, with numerical value
$a_\text{4PN}\simeq 64.6406$, is in perfect agreement (thanks to our
previous adjustment of ambiguity parameters) with GSF numerical
calculations~\cite{vdM16}.

\section{Conclusions}
\label{sec:conclusions}

In this paper, we further investigated the problem of the 4PN
equations of motion of compact binary systems without spins in
harmonic coordinates. In our previous work~\cite{BBBFM16a} (Paper~I),
we computed the Fokker Lagrangian of the motion in harmonic
coordinates or, equivalently, after performing suitable shifts of the
particle's world-lines, the Hamiltonian in ADM-like coordinates.  An
important feature of the dynamics at the 4PN order is that it is
non-local in time, due to the appearance of the tail effect. In the
present paper, we advocated that the infra-red (IR) regularization of
the bound of integrals at spatial infinity in the Einstein-Hilbert
part of the Fokker action is problematic, as different types of
regularizations might lead to different results. However, we
conjectured that the difference is composed of two types of terms
(modulo some irrelevant shifts of the trajectories). Motivated by
this, we introduced two and only two ambiguity parameters reflecting
an incompleteness in our understanding of the IR regularization. Among
these two parameters, one is actually equivalent to the ambiguity
parameter already assumed in Paper~I, so that we actually just added
one extra parameter. The likely existence of a second ambiguity
parameter in harmonic coordinates has been first suggested in
Ref.~\cite{DJS16}. Further work~\cite{BBBFM16c} will be needed to
confirm our conjecture regarding the structure of the ambiguous terms
and to understand whether the use of dimensional regularization
permits determining one combination of these ambiguity parameters.

Next, we obtained the conserved integral of energy and the periastron
advance in the case of circular orbits, fully taking into account the
non-local character of the dynamics brought by the tail term. Thanks
to these two observables, we have been able to uniquely fix the values
of the two ambiguity parameters by requiring that the expressions of
the energy and periastron advance for circular orbits coincide with
those predicted by gravitational self-force (GSF) calculations in the
small mass-ratio limit. Therefore, our 4PN dynamics of compact
binaries, with arbitrary mass ratio, in either Lagrangian or
Hamiltonian form, is complete. However, to reach our goal, we have
postulated a particular structure for the ambiguous part of the Fokker
Lagrangian and relied on it. We leave to future
work~\cite{BBBFM16c} the task of better understanding the nature
of these ambiguities.

An important problem encountered in Paper~I was a discrepancy between
our computation of the conserved energy in the circular orbit limit
for the non-local dynamics and the derivation proposed in
Refs.~\cite{DJS14,DJS16} within the Hamiltonian formalism. The latter
was based on an initial ``localization'' of the Hamiltonian by means
of appropriate non-local shifts of the trajectories, which yielded a
local Hamiltonian in the shifted variables. In the present paper, we
resolved this discrepancy by showing that, when a Hamiltonian is
non-local, there arises an extra term in the conserved integral of
energy (with respect to the value on shell of the Hamiltonian itself)
containing a purely constant (DC type) piece that gives a net
contribution in the case of circular orbits. This extra contribution
was not considered in Paper~I. We thoroughly investigated such extra
terms, both in the integrals of energy and angular momentum, by means
of Fourier series valid for general orbits. Taking into account the
presence of the DC term in the conserved energy for circular orbits,
we recovered the same result as that derived in the Hamiltonian
framework~\cite{DJS14,DJS16}. Finally, after having fixed the two
ambiguity parameters by comparison with GSF results, we found that our
complete 4PN dynamics is in full agreement (for all the terms) with
the results of the ADM Hamiltonian
formalism~\cite{JaraS12,JaraS13,JaraS15,DJS14,DJS16}. However both the
ADM and harmonic gauge methods rely on GSF results to fix the
ambiguity parameters that arise from the IR regularization. In the
future it would be interesting to devise a method that provides an
unambiguous complete result without resorting to GSF.

Future works will extend the equations of motion to the 4.5PN order by
including the dissipative radiation-reaction terms. We will also concentrate
on the computation of the multipole moments of compact binaries and on the
gravitational radiation field to 4.5PN order. A first part of the latter
program, concerning high-order tail effects in the radiation field, has been
recently completed~\cite{MBF16}.

\acknowledgments We thank Gerhard Sch\"afer and Thibault Damour for
informative discussions about the ADM Hamiltonian approach. One of us (L.~Bl.)
is grateful to Leor Barack and Maarten van de Meent for interesting
discussions on the gravitational self-force calculation of the periastron
advance. L.~Be. acknowledges financial support provided under the European
Union's H2020 ERC Consolidator Grant ``Matter and strong-field gravity: New
frontiers in Einstein's theory'' grant agreement no. MaGRaTh–646597. This
project has received funding from the European Union's Horizon 2020 research
and innovation programme under the Marie Sklodowska-Curie grant agreement No
690904.

\appendix

\section{The 4PN Lagrangian in harmonic and ADM coordinates}
\label{app:L}

Our final 4PN Lagrangian in harmonic coordinates is written like in
Paper~I as
\begin{equation}\label{Lstruct}
L = L_\text{N} + \frac{1}{c^2}L_\text{1PN} + \frac{1}{c^4}L_\text{2PN}
+ \frac{1}{c^6}L_\text{3PN} + \frac{1}{c^8}L_\text{4PN} +
\mathcal{O}\left(\frac{1}{c^{10}}\right)\,.
\end{equation}
The terms up to the 3PN order are given by Eqs.~(5.2) in Paper~I, and
the 4PN term is made of a non-local tail piece and many instantaneous
contributions,
\begin{subequations}\label{4PNterm}
\begin{align}
L_\text{4PN} &= L^\text{inst}_\text{4PN} +
L^\text{tail}_\text{4PN}\,,\\ L^\text{inst}_\mathrm{4PN} &=
L^{(0)}_\mathrm{4PN} + G\,L^{(1)}_\mathrm{4PN} +
G^2\,L^{(2)}_\mathrm{4PN} + G^3\,L^{(3)}_\mathrm{4PN} +
G^4\,L^{(4)}_\mathrm{4PN} + G^5\,L^{(5)}_\mathrm{4PN}\,.
\end{align}
\end{subequations}
The tail piece is given by Eq.~(5.4) in Paper~I, while all the
instantaneous terms are provided by Eqs.~(5.6) in Paper~I, except for
the $G^4$ term which is now to be corrected with the extra term in
Eq.~\eqref{Lnew}, with the specific values~\eqref{ambparam} of the
ambiguity parameters. Thus the $G^4$ term in Paper~I is to be replaced
by
\begin{align}\label{G4}
L_\text{4PN}^{(4)}&={} \frac{m_{1}^4 m_{2}}{r_{12}^{3}}
\biggl[\frac{1691807}{25200} (a_1 n_{12}) - \frac{149}{6} (a_2
  n_{12})\biggr] \nonumber\\ & + \frac{m_{1}^3 m_{2}^2}{r_{12}^{3}}
\biggl[\Bigl(- \frac{2470667}{16800} + \frac{1099}{96} \pi^2\Bigr)
  (a_1 n_{12}) + \Bigl(\frac{9246557}{50400} - \frac{555}{64}
  \pi^2\Bigr) (a_2 n_{12})\biggr] \nonumber\\ & + \frac{m_{1}^4
  m_{2}}{r_{12}^{4}}\biggl[\Bigl(\frac{2146}{75} - \frac{880}{3}
  \ln\Bigl[\frac{r_{12}}{r'_{1}}\Bigr]\Bigr) (n_{12} v_1)^2 +
  \Bigl(\frac{3461}{50} + \frac{880}{3}
  \ln\Bigl[\frac{r_{12}}{r'_{1}}\Bigr]\Bigr) (n_{12} v_1) (n_{12} v_2)
  \nonumber\\ &\qquad\qquad - \frac{1165}{12} (n_{12} v_2)^2 + \Bigl(-
  \frac{11479}{300} - \frac{220}{3}
  \ln\Bigl[\frac{r_{12}}{r'_{1}}\Bigr]\Bigr) (v_1 v_2)
  \nonumber\\ &\qquad\qquad + \Bigl(\frac{317}{25} + \frac{220}{3}
  \ln\Bigl[\frac{r_{12}}{r'_{1}}\Bigr]\Bigr) v_1^{2} + \frac{1237}{48}
  v_2^{2}\biggr]\nonumber\\ & + \frac{m_{1}^3 m_{2}^2}{r_{12}^{4}}
\biggl[\Bigl(\frac{26957687}{50400} - \frac{3737}{96} \pi^2 -
  \frac{286}{3} \ln\Bigl[\frac{r_{12}}{r'_{1}}\Bigr]\Bigr) (n_{12}
  v_1)^2 \nonumber\\ &\qquad\qquad + \Bigl(- \frac{4158043}{5040} +
  \frac{179}{4} \pi^2 + 44 \ln\Bigl[\frac{r_{12}}{r'_{1}}\Bigr] + 64
  \ln\Bigl[\frac{r_{12}}{r'_{2}}\Bigr]\Bigr) (n_{12} v_1) (n_{12} v_2)
  \nonumber\\ &\qquad\qquad + \Bigl(\frac{16311923}{50400} -
  \frac{559}{96} \pi^2 + \frac{110}{3}
  \ln\Bigl[\frac{r_{12}}{r'_{1}}\Bigr] - 64
  \ln\Bigl[\frac{r_{12}}{r'_{2}}\Bigr]\Bigr) (n_{12} v_2)^2
  \nonumber\\ &\qquad\qquad + \Bigl( \frac{1568689}{6300} -
  \frac{2627}{192} \pi^2 - \frac{154}{3}
  \ln\Bigl[\frac{r_{12}}{r'_{1}}\Bigr] - 16
  \ln\Bigl[\frac{r_{12}}{r'_{2}}\Bigr]\Bigr) (v_1
  v_2)\nonumber\\ &\qquad\qquad + \Bigl(- \frac{1818001}{12600} +
  \frac{527}{48} \pi^2 + \frac{121}{3}
  \ln\Bigl[\frac{r_{12}}{r'_{1}}\Bigr]\Bigr) v_1^{2}
  \nonumber\\ &\qquad\qquad + \Bigl(- \frac{293443}{3150} +
  \frac{173}{64} \pi^2 + \frac{22}{3}
  \ln\Bigl[\frac{r_{12}}{r'_{1}}\Bigr] + 16
  \ln\Bigl[\frac{r_{12}}{r'_{2}}\Bigr]\Bigr) v_2^{2}\biggr]
\nonumber\\ & + 1 \leftrightarrow 2\,.
\end{align}
Similarly, the $G^4$ instantaneous contribution to the 4PN term in our
Lagrangian in ADM coordinates as given by Eq.~(5.10e) in Paper~I is to
be replaced by
\begin{align}\label{Gbar4}
\tilde{L}_\text{4PN}^{(4)} &={} \frac{m_{1}^4 m_{2}}{r_{12}^{4}}
\biggl[\frac{19341}{1600} (n_{12} v_1)^2 - \frac{15}{8} (v_1 v_2) -
  \frac{16411}{4800} v_1^{2} + \frac{31}{32} v_2^{2}\biggl]
\nonumber\\ & + \frac{m_{1}^3 m_{2}^2}{r_{12}^{4}} \biggl[\Bigl(-
  \frac{6250423}{403200} - \frac{15857}{16384} \pi^2\Bigr) (n_{12}
  v_1)^2 \nonumber\\ &\qquad\qquad + \Bigl(\frac{52572353}{403200} -
  \frac{79385}{24576} \pi^2\Bigr) (n_{12} v_1) (n_{12} v_2)
  \nonumber\\ &\qquad\qquad + \Bigl(- \frac{19635893}{403200}+
  \frac{35603}{24576} \pi^2\Bigr) (n_{12} v_2)^2 +
  \Bigl(\frac{15368099}{403200} - \frac{171041}{24576} \pi^2\Bigr)
  (v_1 v_2) \nonumber\\ &\qquad\qquad + \Bigl(-
  \frac{10602871}{403200} + \frac{193801}{49152} \pi^2\Bigr) v_1^{2}+
  \Bigl(- \frac{5896421}{403200} + \frac{21069}{8192} \pi^2\Bigr)
  v_2^{2}\biggr] \nonumber\\ & + 1 \leftrightarrow 2\,,
\end{align}
with all the other terms in the ADM Lagrangian unchanged with respect
to Paper~I.

\section{Fourier coefficients of the quadrupole moment}
\label{app:fourier}

Here we give the expressions of the Fourier coefficients of the
Newtonian quadrupole moments in terms of combinations of ordinary
Bessel functions. The Fourier coefficients are defined by
Eq.~\eqref{Fourier} and are explicitly given in the Appendix
of~\cite{ABIQ08tail}. We present here some alternative expressions,
based on the decomposition
\begin{equation}\label{Fourierdecompose}
\mathop{{\mathcal{I}}}_{p}{}_{\!\!ij} = \mathop{\mathcal{A}}_{p}
m_0^{i}m_0^{j} + \mathop{\mathcal{B}}_{p}
\,\overline{m}_0^{i}\overline{m}_0^{j} + \mathop{\mathcal{C}}_{p}
\ell^{\langle i}\ell^{j\rangle}\,,
\end{equation}
where we denote $\bm{m}_0=\frac{1}{\sqrt{2}}(\bm{i}+\ui\,\bm{j})$ and
$\overline{\bm{m}}_0=\frac{1}{\sqrt{2}}(\bm{i}-\ui\,\bm{j})$, with
$\bm{i}$, $\bm{j}$ being two fixed orthonormal basis vectors in the
orbital plane, $\bm{i}$ pointing towards the orbit's periastron. Thus,
if $(\bm{n},\bm{\lambda})$ denotes the moving triad in the notation of
Sec.~\ref{sec:integralE}, we have $\bm{m} =
\frac{1}{\sqrt{2}}(\bm{n}+\ui\,\bm{\lambda}) =
\ue^{-\ui\varphi}\,\bm{m}_0$ and $\bm{\ell}=\bm{n}\times\bm{\lambda} =
\ui \,\bm{m}_0\times\overline{\bm{m}}_0$. With this decomposition the
coefficients, functions of the eccentricity $e$ and semi-major axis $a$
of the Keplerian orbit, are relatively simple:
\begin{subequations}\label{IabFourier}
\begin{align}
\mathop{\mathcal{A}}_{p} &= \frac{m\nu\,a^2}{p^2 e^2}\biggl\{ 2 e
J_{p-1}\left(p e\right)\left( p
(1-e^2)+\sqrt{1-e^2}\right)\nonumber\\& \qquad\qquad + J_{p}\left(p
e\right)\left(-2(p+1)\left(1+\sqrt{1-e^2}\right) +
e^2\left[1+2p\left(1+\sqrt{1-e^2}\right)\right]\right)\biggr\}\,,\\
%%%%%%%%%%%%%%%%%%%%%%%%%%%%%%%%%%%%%%%%%%%%%%%%%%%%%%%%%%%%%%%%%
\mathop{\mathcal{B}}_{p} &= \frac{m\nu\,a^2}{p^2 e^2}\biggl\{ 2 e
J_{p-1}\left(p e\right)\left( p
(1-e^2)-\sqrt{1-e^2}\right)\nonumber\\& \qquad\qquad + J_{p}\left(p
e\right)\left(-2(p+1)\left(1-\sqrt{1-e^2}\right) +
e^2\left[1+2p\left(1-\sqrt{1-e^2}\right)\right]\right)\biggr\}\,,\\
%%%%%%%%%%%%%%%%%%%%%%%%%%%%%%%%%%%%%%%%%%%%%%%%%%%%%%%%%%%%%%%%%%
\mathop{\mathcal{C}}_{p} &= \frac{m\nu\,a^2}{p^2} J_{p}\left(p e\right)\,.
\end{align}
\end{subequations}
These expressions are only valid when $p\not= 0$. In the case $p=0$
the coefficients are
\begin{subequations}\label{casp0}
\begin{align}
\mathop{\mathcal{A}}_{0} = \mathop{\mathcal{B}}_{0} =
\frac{5}{4}m\nu\,a^2\,e^2\,,\qquad\mathop{\mathcal{C}}_{0} =
-\frac{1}{2}m\nu\,a^2\left(1+\frac{3}{2}e^2\right)\,.
\end{align}
\end{subequations}
Note that for circular orbits we have (with other modes being zero):
\begin{align}
& \mathop{{\mathcal{I}}}_{2}{}_{\!\!ij} = \frac{m \nu a^2}{2}
  \overline{m}_0^{i}\overline{m}_0^{j}\,, & &
  \mathop{{\mathcal{I}}}_{-2}{}_{\!\!ij} = \frac{m \nu a^2}{2}
  m_0^{i}m_0^{j}\,, & & \mathop{{\mathcal{I}}}_{0}{}_{\!\!ij}
  =-\frac{m \nu a^2}{2} \ell^{\langle i}\ell^{j\rangle}
\end{align}
(\textit{cf.} Footnote~\ref{ft8}).

\section{On Sommerfeld's method of contour integrals}
\label{app:sommerfeld}

In this Appendix, we provide more details on Sommerfeld's method of
contour integrals~\cite{Sommerfeld}, which we used in
Sec.~\ref{sec:periK} to determine the periastron advance for circular
orbits. More precisely we want to show how to adapt the method to the
case of integrals containing a logarithm. The integrals to be
evaluated in our calculation take the form
\begin{equation}\label{periastron_integrals}
\mathcal{I}^{n,p} =\frac{1}{\pi} \int_{r_{p}}^{r_{a}} \frac{\ud
  r}{r^n} \frac{\ln^p r}{\sqrt{R(r,E,h)}}\,,
\end{equation}
for $p=0,1$, where the effective radial potential $R(r,E,h)$ up to 4PN
order has the following structure, extending that in Eq.~(3.4) of
Ref.~\cite{DS88},
\begin{equation}\label{effpot}
R(r,E,h) = A + \frac{2B}{r} + \frac{C}{r^2} + \sum_{i\geqslant 3}
\frac{D_i + E_i \ln r}{r^{i}} \,.
\end{equation}
The coefficients $A$, $B$, $C$, $D_i$ and $E_i$ depend only on the
energy $E$ and angular momentum $h$, and may be considered as mere
constants in the present discussion. While $A$, $B$ and $C$ start at
Newtonian order, $D_i$ and $E_i$ are post-Newtonian expressions. By
construction, the effective potential~\eqref{effpot} vanishes at the
periastron $r=r_{p}$ and at the apastron $r_{a}$, which are the only real
roots of the equation $R=0$. Note that, while logarithmic
contributions start at the 3PN order in our Lagrangian-based
derivation (since the Lagrangian is defined in harmonic coordinates),
they arise only at the 4PN order, and only in the tail part, in our
derivation based on the Hamiltonian in ADM coordinates.

\begin{figure}[t]
\begin{center}
\begin{tabular}{c}
\hspace{-1.2cm}\includegraphics[width=9.0cm,angle=0]{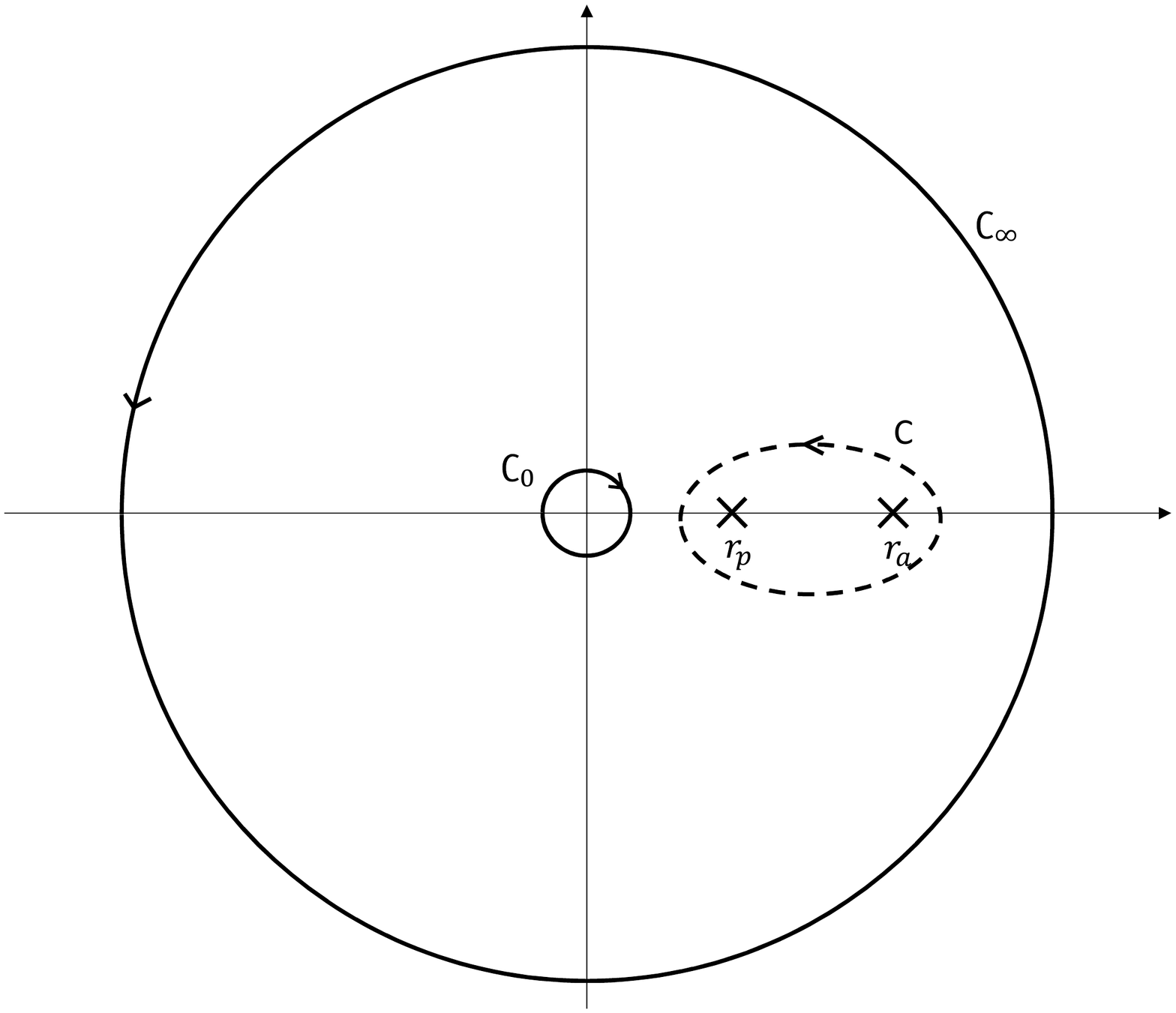}
%\hspace{1.0cm}
\includegraphics[width=9.0cm,angle=0]{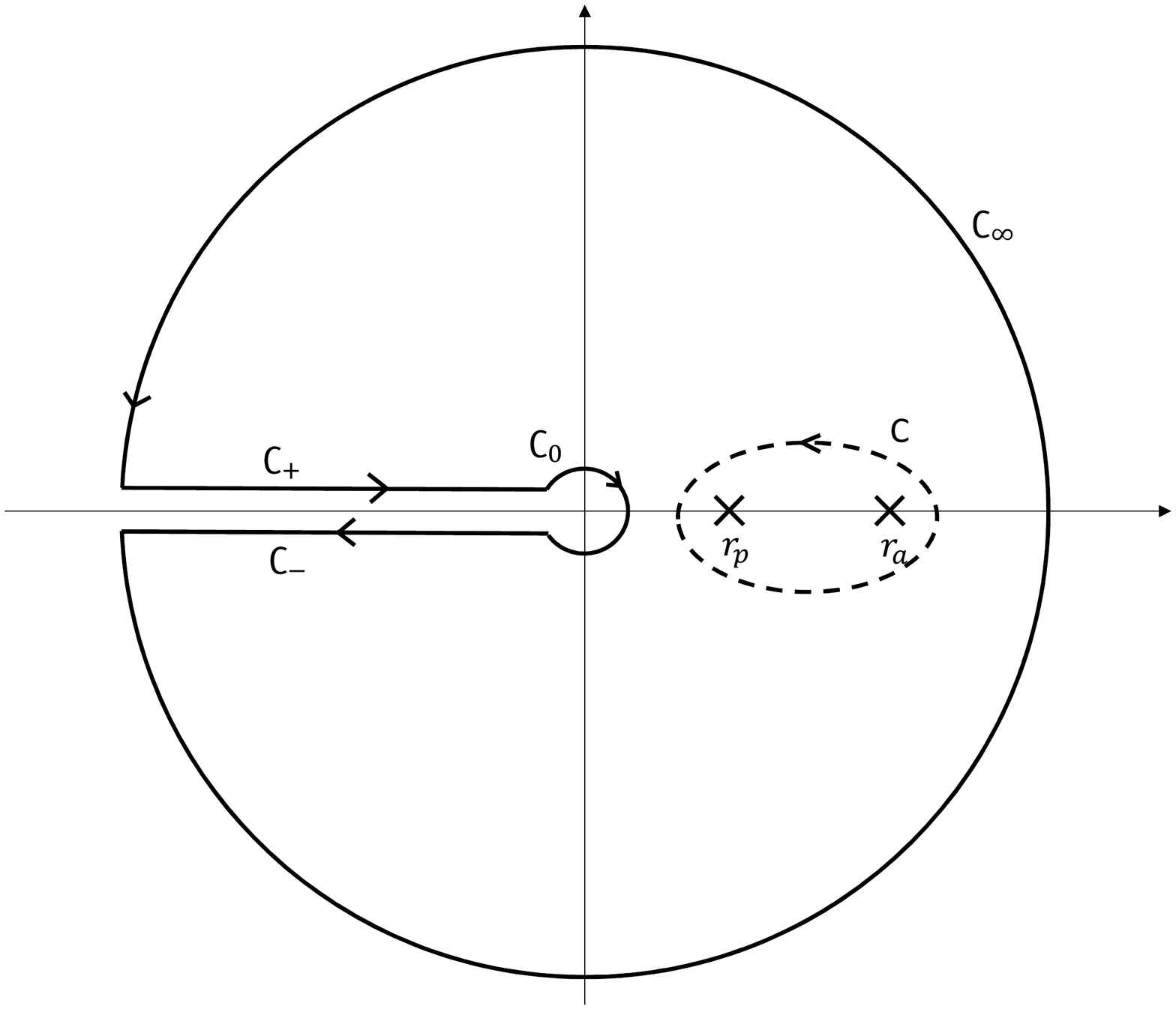}
\end{tabular}
\caption{Left panel: Path of integration of the integrals $I^{n,m}$
  corresponding to the standard, logarithmic-free case. Right panel:
  Path of integration of the integrals $K^{n,m}$ for the logarithmic
  case. The branch cut $\mathbb{R}^{-}$ of the logarithm function is
  avoided.}\label{fig:contour}
\end{center}
\end{figure}

To compute~\eqref{periastron_integrals}, we invoke the procedure
described in Ref.~\cite{DS88}. The first step consists in choosing one
branch cut for $[R(r,E,h)]^{-\frac{1}{2}}$, regarded as a function of
$r$ over the complex plane, to be along the real segment
$[r_{p},r_{a}]$ on the real axis. Since the values of the integrand
over and under that segment are opposite to each other, the original
integral~\eqref{periastron_integrals} is equal to half the
corresponding contour integral over a closed path $\mathrm{C}$ that
surrounds the real segment $[r_{p},r_{a}]$ anti-clockwise (see
Fig.~\ref{fig:contour}):
\begin{equation}\label{periastron_integralsC}
\mathcal{I}^{n,p} = \frac{1}{2\pi}\oint_{\mathrm{C}} \,\frac{\ud r}{r^n}
\frac{\ln^p r}{\sqrt{R(r,E,h)}}\,.
\end{equation}
In a second step, we perform the post-Newtonian expansion of
$[R(r,E,h)]^{-\frac{1}{2}}$ under the integration symbol up to the 4PN
order, holding the contour $\mathrm{C}$ fixed. This yields a linear
combination of elementary integrals of two types, namely
\begin{subequations}\label{intIK}
\begin{align}
I^{n,m} &= \frac{1}{2\pi} \oint_{\mathrm{C}} \,\frac{\ud r}{r^n} \left[
  -1 + \frac{\beta}{r} - \frac{\gamma}{r^2} \right]^{-\frac{m}{2}}
\,,\\ K^{n,m} &= \frac{1}{2\pi} \oint_{\mathrm{C}} \,\frac{\ud r}{r^n}
\ln r \left[ -1 + \frac{\beta}{r} - \frac{\gamma}{r^2}
  \right]^{-\frac{m}{2}} \,.
\end{align}
\end{subequations}
The constants $\beta$ and $\gamma$ are explicitly given by
$\beta=-\frac{Gm^2\nu}{E}$ and $\gamma=-\frac{h^2}{2m\nu E}$, see
Eq.~\eqref{fr}, and both are positive: $\beta,\gamma>0$. For the
integrals $I^{n,m}$, we finally deform the integration path
$\mathrm{C}$ to the contour $\mathrm{C}_0 \cup \mathrm{C}_\infty$
displayed in the left panel of Fig.~\ref{fig:contour}, as in the
method originally devised by Sommerfeld~\cite{Sommerfeld}.
The computation of $I^{n,m}$ then amounts to a straightforward
application of the residue theorem (see Ref.~\cite{DS88} for more
details).

The integrals $K^{n,m}$ have to be handled more carefully. The crucial
point is that we must change the way $\mathrm{C}$ is deformed so as to
avoid the branch cut due to the logarithm of $r$ on the
$\mathbb{R}^{-}$ axis. The final contour in the complex plane is now
replaced by the one shown on the right panel of
Fig.~\ref{fig:contour}. Let us decompose it into the four different
paths $\mathrm{C}_{0}$, $\mathrm{C}_{-}$, $\mathrm{C}_{+}$ and
$\mathrm{C}_{\infty}$ as depicted on the figure. They respectively
correspond to the contours around zero, below the $\mathbb{R}^{-}$
branch cut, above the $\mathbb{R}^{-}$ branch cut, and at
infinity. This decomposition induces the corresponding splitting
\begin{equation}
K^{n,m} = K_{0}^{n,m} + K_{-}^{n,m} + K_{+}^{n,m} + K_{\infty}^{n,m}
\,.
\end{equation}
On $\mathrm{C}_{0}$, we can write $r=\eta\,\mathrm{e}^{-\ui \,
  \theta}$ with $\eta$ a small positive number and $\theta$ going from
$\pi$ to $-\pi$,\footnote{Strictly speaking, we should first integrate
  with $\theta$ going from $\pi-\epsilon$ to $-\pi+\epsilon$, and then
  take the limit $\epsilon\rightarrow 0$ in the end. We have checked
  that the end result comes out the same.} so that the logarithm
becomes $\ln r=\ln \eta-\ui \, \theta$.  We are led to
\begin{equation}
K_{0}^{n,m} = \frac{\ui \, (-)^{m/2}}{2\pi
  \,\gamma^{m/2}\,\eta^{n-m-1}} \int_{-\pi}^{\pi} \, \ud\theta\,
\ue^{\ui\, (n-m-1)\theta} \left( \ln\eta-\ui\, \theta \right) \left[ 1
  - \frac{\beta}{\gamma} \eta\, \ue^{-\ui\, \theta} +\frac{1}{\gamma}
  \,\eta^{2} \ue^{-2\ui\, \theta} \right]^{-\frac{m}{2}} \,.
\end{equation}
After expanding the above expression in powers of $\eta$, the
resulting integrals are immediate to evaluate. For example, when $n=5$
and $m=3$, we obtain
\begin{equation}
K_{0}^{5,3} = -\frac{1}{\gamma^{3/2}\,\eta} -
\frac{3\beta}{2\gamma^{5/2}} \ln\eta \,.
\end{equation}
As for the integrals at infinity that we have to consider, they are
all zero:
\begin{equation}
K_{\infty}^{n,m} = 0 \,.
\end{equation}
We are left with the two integrals under and over the branch cut
$\mathbb{R}^{-}$. On the contour $\mathrm{C}_{-}$, we use the
parametrization $r=\rho\,\ue^{-\ui \pi}$, for $\rho$ going from $\eta$
to some radius $R$, with $R\gg 1$ and $\eta\ll 1$. Similarly, on the
contour $\mathrm{C}_{+}$ we take $r=\rho\,\ue^{\ui \pi}$, for $\rho$
going from $R$ to $\eta$. The difference of $2\pi$ is due to the
presence of the branch cut for the complex logarithm.  Adding the two
integrals, we get
\begin{equation}
K_{-}^{n,m} + K_{+}^{n,m} = (-)^n\, \ui \int_{\eta}^{R} \,\frac{\ud
  \rho}{\rho^n} \left[ -1-\frac{\beta}{\rho}-\frac{\gamma}{\rho^2}
  \right]^{-\frac{m}{2}} \,.
\end{equation}
As one can see, the logarithms have canceled out. In the end, there
remains a well-known, tabulated integral~\cite{GR}. For example, for
$n=5$ and $m=1$, we find
\begin{align}\nonumber
K_{-}^{5,1} + K_{+}^{5,1} = & -\frac{1}{3\sqrt{\gamma}\eta^3} +
\frac{\beta}{4\gamma^{3/2}\eta^2} +
\left(-\frac{3\beta^2}{8\gamma^{5/2}} +
\frac{1}{2\gamma^{3/2}}\right)\frac{1}{\eta} -
\frac{37\beta^3}{96\gamma^{7/2}} + \frac{5\beta^2}{8\gamma^3} \\ & +
\frac{5\beta}{8\gamma^{5/2}} - \frac{2}{3\gamma^2} + \left(
-\frac{5\beta^3}{16\gamma^{7/2}} + \frac{3\beta}{4\gamma^{5/2}}
\right) \left[ \ln\left( \frac{\beta}{4\gamma} +
  \frac{1}{2\sqrt{\gamma}} \right) + \ln\eta\right] \,.
\end{align}
Finally, it remains to add all the different pieces and take the limits
$\eta \rightarrow 0$ and $R \rightarrow +\infty$.

\bibliography{ListeRef}

\begin{thebibliography}{57}
\expandafter\ifx\csname natexlab\endcsname\relax\def\natexlab#1{#1}\fi
\expandafter\ifx\csname bibnamefont\endcsname\relax
  \def\bibnamefont#1{#1}\fi
\expandafter\ifx\csname bibfnamefont\endcsname\relax
  \def\bibfnamefont#1{#1}\fi
\expandafter\ifx\csname citenamefont\endcsname\relax
  \def\citenamefont#1{#1}\fi
\expandafter\ifx\csname url\endcsname\relax
  \def\url#1{\texttt{#1}}\fi
\expandafter\ifx\csname urlprefix\endcsname\relax\def\urlprefix{URL }\fi
\providecommand{\bibinfo}[2]{#2}
\providecommand{\eprint}[2][]{\url{#2}}

\bibitem[{\citenamefont{Bernard
  et~al.}(2016{\natexlab{a}})\citenamefont{Bernard, Blanchet, Boh\'e, Faye, and
  Marsat}}]{BBBFM16a}
\bibinfo{author}{\bibfnamefont{L.}~\bibnamefont{Bernard}},
  \bibinfo{author}{\bibfnamefont{L.}~\bibnamefont{Blanchet}},
  \bibinfo{author}{\bibfnamefont{A.}~\bibnamefont{Boh\'e}},
  \bibinfo{author}{\bibfnamefont{G.}~\bibnamefont{Faye}}, \bibnamefont{and}
  \bibinfo{author}{\bibfnamefont{S.}~\bibnamefont{Marsat}},
  \bibinfo{journal}{Phys. Rev. D} \textbf{\bibinfo{volume}{93}},
  \bibinfo{pages}{084037} (\bibinfo{year}{2016}{\natexlab{a}}),
  \eprint{arXiv:1512.02876 [gr-qc]}.

\bibitem[{\citenamefont{Blanchet}(2014)}]{Bliving14}
\bibinfo{author}{\bibfnamefont{L.}~\bibnamefont{Blanchet}},
  \bibinfo{journal}{Living Rev. Rel.} \textbf{\bibinfo{volume}{17}},
  \bibinfo{pages}{2} (\bibinfo{year}{2014}), \eprint{arXiv:1310.1528 [gr-qc]}.

\bibitem[{\citenamefont{Jaranowski and Sch\"afer}(1998)}]{JaraS98}
\bibinfo{author}{\bibfnamefont{P.}~\bibnamefont{Jaranowski}} \bibnamefont{and}
  \bibinfo{author}{\bibfnamefont{G.}~\bibnamefont{Sch\"afer}},
  \bibinfo{journal}{Phys. Rev. D} \textbf{\bibinfo{volume}{57}},
  \bibinfo{pages}{7274} (\bibinfo{year}{1998}), \eprint{gr-qc/9712075}.

\bibitem[{\citenamefont{Jaranowski and Sch\"afer}(1999)}]{JaraS99}
\bibinfo{author}{\bibfnamefont{P.}~\bibnamefont{Jaranowski}} \bibnamefont{and}
  \bibinfo{author}{\bibfnamefont{G.}~\bibnamefont{Sch\"afer}},
  \bibinfo{journal}{Phys. Rev. D} \textbf{\bibinfo{volume}{60}},
  \bibinfo{pages}{124003} (\bibinfo{year}{1999}), \eprint{gr-qc/9906092}.

\bibitem[{\citenamefont{Jaranowski and Sch\"afer}(2000)}]{JaraS00}
\bibinfo{author}{\bibfnamefont{P.}~\bibnamefont{Jaranowski}} \bibnamefont{and}
  \bibinfo{author}{\bibfnamefont{G.}~\bibnamefont{Sch\"afer}},
  \bibinfo{journal}{Ann. Phys. (Berlin)} \textbf{\bibinfo{volume}{9}},
  \bibinfo{pages}{378} (\bibinfo{year}{2000}), \eprint{gr-qc/0003054}.

\bibitem[{\citenamefont{Damour et~al.}(2000)\citenamefont{Damour, Jaranowski,
  and Sch\"afer}}]{DJSpoinc}
\bibinfo{author}{\bibfnamefont{T.}~\bibnamefont{Damour}},
  \bibinfo{author}{\bibfnamefont{P.}~\bibnamefont{Jaranowski}},
  \bibnamefont{and}
  \bibinfo{author}{\bibfnamefont{G.}~\bibnamefont{Sch\"afer}},
  \bibinfo{journal}{Phys. Rev. D} \textbf{\bibinfo{volume}{62}},
  \bibinfo{pages}{021501(R)} (\bibinfo{year}{2000}), \bibinfo{note}{erratum
  Phys. Rev. D {\bf 63}, 029903(E) (2000)}, \eprint{gr-qc/0003051}.

\bibitem[{\citenamefont{Damour et~al.}(2001{\natexlab{a}})\citenamefont{Damour,
  Jaranowski, and Sch\"afer}}]{DJSequiv}
\bibinfo{author}{\bibfnamefont{T.}~\bibnamefont{Damour}},
  \bibinfo{author}{\bibfnamefont{P.}~\bibnamefont{Jaranowski}},
  \bibnamefont{and}
  \bibinfo{author}{\bibfnamefont{G.}~\bibnamefont{Sch\"afer}},
  \bibinfo{journal}{Phys. Rev. D} \textbf{\bibinfo{volume}{63}},
  \bibinfo{pages}{044021} (\bibinfo{year}{2001}{\natexlab{a}}),
  \bibinfo{note}{erratum Phys. Rev. D {\bf 66}, 029901(E) (2002)},
  \eprint{gr-qc/0010040}.

\bibitem[{\citenamefont{Damour et~al.}(2001{\natexlab{b}})\citenamefont{Damour,
  Jaranowski, and Sch\"afer}}]{DJSdim}
\bibinfo{author}{\bibfnamefont{T.}~\bibnamefont{Damour}},
  \bibinfo{author}{\bibfnamefont{P.}~\bibnamefont{Jaranowski}},
  \bibnamefont{and}
  \bibinfo{author}{\bibfnamefont{G.}~\bibnamefont{Sch\"afer}},
  \bibinfo{journal}{Phys. Lett. B} \textbf{\bibinfo{volume}{513}},
  \bibinfo{pages}{147} (\bibinfo{year}{2001}{\natexlab{b}}),
  \eprint{gr-qc/0105038}.

\bibitem[{\citenamefont{Blanchet and Faye}(2000{\natexlab{a}})}]{BF00}
\bibinfo{author}{\bibfnamefont{L.}~\bibnamefont{Blanchet}} \bibnamefont{and}
  \bibinfo{author}{\bibfnamefont{G.}~\bibnamefont{Faye}},
  \bibinfo{journal}{Phys. Lett. A} \textbf{\bibinfo{volume}{271}},
  \bibinfo{pages}{58} (\bibinfo{year}{2000}{\natexlab{a}}),
  \eprint{gr-qc/0004009}.

\bibitem[{\citenamefont{Blanchet and Faye}(2001)}]{BFeom}
\bibinfo{author}{\bibfnamefont{L.}~\bibnamefont{Blanchet}} \bibnamefont{and}
  \bibinfo{author}{\bibfnamefont{G.}~\bibnamefont{Faye}},
  \bibinfo{journal}{Phys. Rev. D} \textbf{\bibinfo{volume}{63}},
  \bibinfo{pages}{062005} (\bibinfo{year}{2001}), \eprint{gr-qc/0007051}.

\bibitem[{\citenamefont{Blanchet and Faye}(2000{\natexlab{b}})}]{BFreg}
\bibinfo{author}{\bibfnamefont{L.}~\bibnamefont{Blanchet}} \bibnamefont{and}
  \bibinfo{author}{\bibfnamefont{G.}~\bibnamefont{Faye}}, \bibinfo{journal}{J.
  Math. Phys.} \textbf{\bibinfo{volume}{41}}, \bibinfo{pages}{7675}
  (\bibinfo{year}{2000}{\natexlab{b}}), \eprint{gr-qc/0004008}.

\bibitem[{\citenamefont{de~Andrade et~al.}(2001)\citenamefont{de~Andrade,
  Blanchet, and Faye}}]{ABF01}
\bibinfo{author}{\bibfnamefont{V.}~\bibnamefont{de~Andrade}},
  \bibinfo{author}{\bibfnamefont{L.}~\bibnamefont{Blanchet}}, \bibnamefont{and}
  \bibinfo{author}{\bibfnamefont{G.}~\bibnamefont{Faye}},
  \bibinfo{journal}{Class. Quant. Grav.} \textbf{\bibinfo{volume}{18}},
  \bibinfo{pages}{753} (\bibinfo{year}{2001}), \eprint{gr-qc/0011063}.

\bibitem[{\citenamefont{Blanchet and Iyer}(2003)}]{BI03CM}
\bibinfo{author}{\bibfnamefont{L.}~\bibnamefont{Blanchet}} \bibnamefont{and}
  \bibinfo{author}{\bibfnamefont{B.~R.} \bibnamefont{Iyer}},
  \bibinfo{journal}{Class. Quant. Grav.} \textbf{\bibinfo{volume}{20}},
  \bibinfo{pages}{755} (\bibinfo{year}{2003}), \eprint{gr-qc/0209089}.

\bibitem[{\citenamefont{Blanchet et~al.}(2004)\citenamefont{Blanchet, Damour,
  and Esposito-Far{\`e}se}}]{BDE04}
\bibinfo{author}{\bibfnamefont{L.}~\bibnamefont{Blanchet}},
  \bibinfo{author}{\bibfnamefont{T.}~\bibnamefont{Damour}}, \bibnamefont{and}
  \bibinfo{author}{\bibfnamefont{G.}~\bibnamefont{Esposito-Far{\`e}se}},
  \bibinfo{journal}{Phys. Rev. D} \textbf{\bibinfo{volume}{69}},
  \bibinfo{pages}{124007} (\bibinfo{year}{2004}), \eprint{gr-qc/0311052}.

\bibitem[{\citenamefont{Itoh et~al.}(2000)\citenamefont{Itoh, Futamase, and
  Asada}}]{IFA00}
\bibinfo{author}{\bibfnamefont{Y.~.} \bibnamefont{Itoh}},
  \bibinfo{author}{\bibfnamefont{T.}~\bibnamefont{Futamase}}, \bibnamefont{and}
  \bibinfo{author}{\bibfnamefont{H.}~\bibnamefont{Asada}},
  \bibinfo{journal}{Phys. Rev. D} \textbf{\bibinfo{volume}{62}},
  \bibinfo{pages}{064002} (\bibinfo{year}{2000}), \eprint{gr-qc/9910052}.

\bibitem[{\citenamefont{Itoh et~al.}(2001)\citenamefont{Itoh, Futamase, and
  Asada}}]{IFA01}
\bibinfo{author}{\bibfnamefont{Y.}~\bibnamefont{Itoh}},
  \bibinfo{author}{\bibfnamefont{T.}~\bibnamefont{Futamase}}, \bibnamefont{and}
  \bibinfo{author}{\bibfnamefont{H.}~\bibnamefont{Asada}},
  \bibinfo{journal}{Phys. Rev. D} \textbf{\bibinfo{volume}{63}},
  \bibinfo{pages}{064038} (\bibinfo{year}{2001}), \eprint{gr-qc/0101114}.

\bibitem[{\citenamefont{Itoh and Futamase}(2003)}]{itoh1}
\bibinfo{author}{\bibfnamefont{Y.}~\bibnamefont{Itoh}} \bibnamefont{and}
  \bibinfo{author}{\bibfnamefont{T.}~\bibnamefont{Futamase}},
  \bibinfo{journal}{Phys. Rev. D} \textbf{\bibinfo{volume}{68}},
  \bibinfo{pages}{121501(R)} (\bibinfo{year}{2003}), \eprint{gr-qc/0310028}.

\bibitem[{\citenamefont{Itoh}(2004)}]{itoh2}
\bibinfo{author}{\bibfnamefont{Y.}~\bibnamefont{Itoh}}, \bibinfo{journal}{Phys.
  Rev. D} \textbf{\bibinfo{volume}{69}}, \bibinfo{pages}{064018}
  (\bibinfo{year}{2004}), \eprint{gr-qc/0310029}.

\bibitem[{\citenamefont{Futamase and Itoh}(2007)}]{ItohLR}
\bibinfo{author}{\bibfnamefont{T.}~\bibnamefont{Futamase}} \bibnamefont{and}
  \bibinfo{author}{\bibfnamefont{T.}~\bibnamefont{Itoh}},
  \bibinfo{journal}{Living Rev. Rel.} \textbf{\bibinfo{volume}{10}},
  \bibinfo{pages}{2} (\bibinfo{year}{2007}).

\bibitem[{\citenamefont{Foffa and Sturani}(2011)}]{FS3PN}
\bibinfo{author}{\bibfnamefont{S.}~\bibnamefont{Foffa}} \bibnamefont{and}
  \bibinfo{author}{\bibfnamefont{R.}~\bibnamefont{Sturani}},
  \bibinfo{journal}{Phys. Rev. D} \textbf{\bibinfo{volume}{84}},
  \bibinfo{pages}{044031} (\bibinfo{year}{2011}), \eprint{arXiv:1104.1122
  [gr-qc]}.

\bibitem[{\citenamefont{Jaranowski and Sch{\"a}fer}(2012)}]{JaraS12}
\bibinfo{author}{\bibfnamefont{P.}~\bibnamefont{Jaranowski}} \bibnamefont{and}
  \bibinfo{author}{\bibfnamefont{G.}~\bibnamefont{Sch{\"a}fer}},
  \bibinfo{journal}{Phys. Rev. D} \textbf{\bibinfo{volume}{86}},
  \bibinfo{pages}{061503(R)} (\bibinfo{year}{2012}), \eprint{arXiv:1207.5448
  [gr-qc]}.

\bibitem[{\citenamefont{Jaranowski and Sch{\"a}fer}(2013)}]{JaraS13}
\bibinfo{author}{\bibfnamefont{P.}~\bibnamefont{Jaranowski}} \bibnamefont{and}
  \bibinfo{author}{\bibfnamefont{G.}~\bibnamefont{Sch{\"a}fer}},
  \bibinfo{journal}{Phys. Rev. D} \textbf{\bibinfo{volume}{87}},
  \bibinfo{pages}{081503(R)} (\bibinfo{year}{2013}), \eprint{arXiv:1303.3225
  [gr-qc]}.

\bibitem[{\citenamefont{Jaranowski and Sch{\"a}fer}(2015)}]{JaraS15}
\bibinfo{author}{\bibfnamefont{P.}~\bibnamefont{Jaranowski}} \bibnamefont{and}
  \bibinfo{author}{\bibfnamefont{G.}~\bibnamefont{Sch{\"a}fer}},
  \bibinfo{journal}{Phys. Rev. D} \textbf{\bibinfo{volume}{92}},
  \bibinfo{pages}{124043} (\bibinfo{year}{2015}), \eprint{arXiv:1508.01016
  [gr-qc]}.

\bibitem[{\citenamefont{Foffa and Sturani}(2012)}]{FS4PN}
\bibinfo{author}{\bibfnamefont{S.}~\bibnamefont{Foffa}} \bibnamefont{and}
  \bibinfo{author}{\bibfnamefont{R.}~\bibnamefont{Sturani}},
  \bibinfo{journal}{Phys. Rev. D} \textbf{\bibinfo{volume}{87}},
  \bibinfo{pages}{064011} (\bibinfo{year}{2012}), \eprint{arXiv:1206.7087
  [gr-qc]}.

\bibitem[{\citenamefont{Damour et~al.}(2014)\citenamefont{Damour, Jaranowski,
  and Sch{\"a}fer}}]{DJS14}
\bibinfo{author}{\bibfnamefont{T.}~\bibnamefont{Damour}},
  \bibinfo{author}{\bibfnamefont{P.}~\bibnamefont{Jaranowski}},
  \bibnamefont{and}
  \bibinfo{author}{\bibfnamefont{G.}~\bibnamefont{Sch{\"a}fer}},
  \bibinfo{journal}{Phys. Rev. D} \textbf{\bibinfo{volume}{89}},
  \bibinfo{pages}{064058} (\bibinfo{year}{2014}), \eprint{arXiv:1401.4548
  [gr-qc]}.

\bibitem[{\citenamefont{Damour et~al.}(2016)\citenamefont{Damour, Jaranowski,
  and Sch{\"a}fer}}]{DJS16}
\bibinfo{author}{\bibfnamefont{T.}~\bibnamefont{Damour}},
  \bibinfo{author}{\bibfnamefont{P.}~\bibnamefont{Jaranowski}},
  \bibnamefont{and}
  \bibinfo{author}{\bibfnamefont{G.}~\bibnamefont{Sch{\"a}fer}},
  \bibinfo{journal}{Phys. Rev. D} \textbf{\bibinfo{volume}{93}},
  \bibinfo{pages}{084014} (\bibinfo{year}{2016}), \eprint{arXiv:1601.01283
  [gr-qc]}.

\bibitem[{\citenamefont{Blanchet and Damour}(1988)}]{BD88}
\bibinfo{author}{\bibfnamefont{L.}~\bibnamefont{Blanchet}} \bibnamefont{and}
  \bibinfo{author}{\bibfnamefont{T.}~\bibnamefont{Damour}},
  \bibinfo{journal}{Phys. Rev. D} \textbf{\bibinfo{volume}{37}},
  \bibinfo{pages}{1410} (\bibinfo{year}{1988}).

\bibitem[{\citenamefont{Blanchet}(1993)}]{B93}
\bibinfo{author}{\bibfnamefont{L.}~\bibnamefont{Blanchet}},
  \bibinfo{journal}{Phys. Rev. D} \textbf{\bibinfo{volume}{47}},
  \bibinfo{pages}{4392} (\bibinfo{year}{1993}).

\bibitem[{\citenamefont{Detweiler}(2011)}]{Detweilerorleans}
\bibinfo{author}{\bibfnamefont{S.}~\bibnamefont{Detweiler}}, in
  \emph{\bibinfo{booktitle}{Mass and motion in general relativity}}, edited by
  \bibinfo{editor}{\bibfnamefont{L.}~\bibnamefont{Blanchet}},
  \bibinfo{editor}{\bibfnamefont{A.}~\bibnamefont{Spallicci}},
  \bibnamefont{and} \bibinfo{editor}{\bibfnamefont{B.}~\bibnamefont{Whiting}}
  (\bibinfo{publisher}{Springer}, \bibinfo{year}{2011}), p.
  \bibinfo{pages}{271}.

\bibitem[{\citenamefont{Barack}(2011)}]{Barackorleans}
\bibinfo{author}{\bibfnamefont{L.}~\bibnamefont{Barack}}, in
  \emph{\bibinfo{booktitle}{Mass and motion in general relativity}}, edited by
  \bibinfo{editor}{\bibfnamefont{L.}~\bibnamefont{Blanchet}},
  \bibinfo{editor}{\bibfnamefont{A.}~\bibnamefont{Spallicci}},
  \bibnamefont{and} \bibinfo{editor}{\bibfnamefont{B.}~\bibnamefont{Whiting}}
  (\bibinfo{publisher}{Springer}, \bibinfo{year}{2011}), p.
  \bibinfo{pages}{327}.

\bibitem[{\citenamefont{Blanchet et~al.}(2010)\citenamefont{Blanchet,
  Detweiler, Le~Tiec, and Whiting}}]{BDLW10b}
\bibinfo{author}{\bibfnamefont{L.}~\bibnamefont{Blanchet}},
  \bibinfo{author}{\bibfnamefont{S.}~\bibnamefont{Detweiler}},
  \bibinfo{author}{\bibfnamefont{A.}~\bibnamefont{Le~Tiec}}, \bibnamefont{and}
  \bibinfo{author}{\bibfnamefont{B.}~\bibnamefont{Whiting}},
  \bibinfo{journal}{Phys. Rev. D} \textbf{\bibinfo{volume}{81}},
  \bibinfo{pages}{084033} (\bibinfo{year}{2010}), \eprint{arXiv:1002.0726
  [gr-qc]}.

\bibitem[{\citenamefont{Le~Tiec
  et~al.}(2012{\natexlab{a}})\citenamefont{Le~Tiec, Blanchet, and
  Whiting}}]{LBW12}
\bibinfo{author}{\bibfnamefont{A.}~\bibnamefont{Le~Tiec}},
  \bibinfo{author}{\bibfnamefont{L.}~\bibnamefont{Blanchet}}, \bibnamefont{and}
  \bibinfo{author}{\bibfnamefont{B.}~\bibnamefont{Whiting}},
  \bibinfo{journal}{Phys. Rev. D} \textbf{\bibinfo{volume}{85}},
  \bibinfo{pages}{064039} (\bibinfo{year}{2012}{\natexlab{a}}),
  \eprint{arXiv:1111.5378 [gr-qc]}.

\bibitem[{\citenamefont{Le~Tiec
  et~al.}(2012{\natexlab{b}})\citenamefont{Le~Tiec, Barausse, and
  Buonanno}}]{LBB12}
\bibinfo{author}{\bibfnamefont{A.}~\bibnamefont{Le~Tiec}},
  \bibinfo{author}{\bibfnamefont{E.}~\bibnamefont{Barausse}}, \bibnamefont{and}
  \bibinfo{author}{\bibfnamefont{A.}~\bibnamefont{Buonanno}},
  \bibinfo{journal}{Phys. Rev. Lett.} \textbf{\bibinfo{volume}{108}},
  \bibinfo{pages}{131103} (\bibinfo{year}{2012}{\natexlab{b}}),
  \eprint{arXiv:1111.5609 [gr-qc]}.

\bibitem[{\citenamefont{Bini and Damour}(2013)}]{BiniD13}
\bibinfo{author}{\bibfnamefont{D.}~\bibnamefont{Bini}} \bibnamefont{and}
  \bibinfo{author}{\bibfnamefont{T.}~\bibnamefont{Damour}},
  \bibinfo{journal}{Phys. Rev. D} \textbf{\bibinfo{volume}{87}},
  \bibinfo{pages}{121501(R)} (\bibinfo{year}{2013}), \eprint{arXiv:1305.4884
  [gr-qc]}.

\bibitem[{\citenamefont{Barack et~al.}(2010)\citenamefont{Barack, Damour, and
  Sago}}]{BDS10}
\bibinfo{author}{\bibfnamefont{L.}~\bibnamefont{Barack}},
  \bibinfo{author}{\bibfnamefont{T.}~\bibnamefont{Damour}}, \bibnamefont{and}
  \bibinfo{author}{\bibfnamefont{N.}~\bibnamefont{Sago}},
  \bibinfo{journal}{Phys. Rev. D} \textbf{\bibinfo{volume}{82}},
  \bibinfo{pages}{084036} (\bibinfo{year}{2010}), \eprint{arXiv:1008.0935
  [gr-qc]}.

\bibitem[{\citenamefont{Le~Tiec et~al.}(2011)\citenamefont{Le~Tiec, Mrou\'e,
  Barack, Buonanno, Pfeiffer, Sago, and Taracchini}}]{Letal11}
\bibinfo{author}{\bibfnamefont{A.}~\bibnamefont{Le~Tiec}},
  \bibinfo{author}{\bibfnamefont{A.}~\bibnamefont{Mrou\'e}},
  \bibinfo{author}{\bibfnamefont{L.}~\bibnamefont{Barack}},
  \bibinfo{author}{\bibfnamefont{A.}~\bibnamefont{Buonanno}},
  \bibinfo{author}{\bibfnamefont{H.}~\bibnamefont{Pfeiffer}},
  \bibinfo{author}{\bibfnamefont{N.}~\bibnamefont{Sago}}, \bibnamefont{and}
  \bibinfo{author}{\bibfnamefont{A.}~\bibnamefont{Taracchini}},
  \bibinfo{journal}{Phys. Rev. Lett.} \textbf{\bibinfo{volume}{107}},
  \bibinfo{pages}{141101} (\bibinfo{year}{2011}), \eprint{arXiv:1106.3278
  [gr-qc]}.

\bibitem[{\citenamefont{van~de Meent}(2016)}]{vdM16}
\bibinfo{author}{\bibfnamefont{M.}~\bibnamefont{van~de Meent}}
  (\bibinfo{year}{2016}), \eprint{1610.03497}.

\bibitem[{\citenamefont{Damour}(2010)}]{D10sf}
\bibinfo{author}{\bibfnamefont{T.}~\bibnamefont{Damour}},
  \bibinfo{journal}{Phys. Rev. D} \textbf{\bibinfo{volume}{81}},
  \bibinfo{pages}{024017} (\bibinfo{year}{2010}), \eprint{arXiv:0910.5533
  [gr-qc]}.

\bibitem[{\citenamefont{Damour et~al.}(2015)\citenamefont{Damour, Jaranowski,
  and Sch{\"a}fer}}]{DJS15eob}
\bibinfo{author}{\bibfnamefont{T.}~\bibnamefont{Damour}},
  \bibinfo{author}{\bibfnamefont{P.}~\bibnamefont{Jaranowski}},
  \bibnamefont{and}
  \bibinfo{author}{\bibfnamefont{G.}~\bibnamefont{Sch{\"a}fer}},
  \bibinfo{journal}{Phys. Rev. D} \textbf{\bibinfo{volume}{91}},
  \bibinfo{pages}{084024} (\bibinfo{year}{2015}), \eprint{arXiv:1502.07245
  [gr-qc]}.

\bibitem[{\citenamefont{Le~Tiec}(2015)}]{L15}
\bibinfo{author}{\bibfnamefont{A.}~\bibnamefont{Le~Tiec}},
  \bibinfo{journal}{Phys. Rev. D} \textbf{\bibinfo{volume}{92}},
  \bibinfo{pages}{084021} (\bibinfo{year}{2015}), \eprint{arXiv:1506.05648
  [gr-qc]}.

\bibitem[{\citenamefont{Blanchet and Le~Tiec}(2016)}]{BL16}
\bibinfo{author}{\bibfnamefont{L.}~\bibnamefont{Blanchet}} \bibnamefont{and}
  \bibinfo{author}{\bibfnamefont{A.}~\bibnamefont{Le~Tiec}}
  (\bibinfo{year}{2016}), \bibinfo{note}{in preparation}.

\bibitem[{\citenamefont{Blanchet et~al.}(2002)\citenamefont{Blanchet, Iyer, and
  Joguet}}]{BIJ02}
\bibinfo{author}{\bibfnamefont{L.}~\bibnamefont{Blanchet}},
  \bibinfo{author}{\bibfnamefont{B.~R.} \bibnamefont{Iyer}}, \bibnamefont{and}
  \bibinfo{author}{\bibfnamefont{B.}~\bibnamefont{Joguet}},
  \bibinfo{journal}{Phys. Rev. D} \textbf{\bibinfo{volume}{65}},
  \bibinfo{pages}{064005} (\bibinfo{year}{2002}), \bibinfo{note}{erratum
  \textit{Phys. Rev. D}, 71:129903(E), 2005}, \eprint{gr-qc/0105098}.

\bibitem[{\citenamefont{Bernard
  et~al.}(2016{\natexlab{b}})\citenamefont{Bernard, Blanchet, Boh\'e, Faye, and
  Marsat}}]{BBBFM16c}
\bibinfo{author}{\bibfnamefont{L.}~\bibnamefont{Bernard}},
  \bibinfo{author}{\bibfnamefont{L.}~\bibnamefont{Blanchet}},
  \bibinfo{author}{\bibfnamefont{A.}~\bibnamefont{Boh\'e}},
  \bibinfo{author}{\bibfnamefont{G.}~\bibnamefont{Faye}}, \bibnamefont{and}
  \bibinfo{author}{\bibfnamefont{S.}~\bibnamefont{Marsat}}
  (\bibinfo{year}{2016}{\natexlab{b}}), \bibinfo{note}{in progress}.

\bibitem[{\citenamefont{Hadamard}(1932)}]{hadamard}
\bibinfo{author}{\bibfnamefont{J.}~\bibnamefont{Hadamard}},
  \emph{\bibinfo{title}{Le probl\`eme de Cauchy et les \'equations aux
  d\'eriv\'ees partielles lin\'eaires hyperboliques}}
  (\bibinfo{publisher}{Hermann}, \bibinfo{address}{Paris},
  \bibinfo{year}{1932}).

\bibitem[{\citenamefont{Schwartz}(1978)}]{schwartz}
\bibinfo{author}{\bibfnamefont{L.}~\bibnamefont{Schwartz}},
  \emph{\bibinfo{title}{Th\'eorie des distributions}}
  (\bibinfo{publisher}{Hermann}, \bibinfo{address}{Paris},
  \bibinfo{year}{1978}).

\bibitem[{\citenamefont{Llosa and Vives}(1994)}]{Llosa}
\bibinfo{author}{\bibfnamefont{J.}~\bibnamefont{Llosa}} \bibnamefont{and}
  \bibinfo{author}{\bibfnamefont{J.}~\bibnamefont{Vives}}, \bibinfo{journal}{J.
  Math. Phys.} \textbf{\bibinfo{volume}{35}}, \bibinfo{pages}{2856}
  (\bibinfo{year}{1994}).

\bibitem[{\citenamefont{Ferialdi and Bassi}(2012)}]{Ferialdi}
\bibinfo{author}{\bibfnamefont{L.}~\bibnamefont{Ferialdi}} \bibnamefont{and}
  \bibinfo{author}{\bibfnamefont{A.}~\bibnamefont{Bassi}},
  \bibinfo{journal}{Eur. Phys. Lett.} \textbf{\bibinfo{volume}{98}},
  \bibinfo{pages}{30009} (\bibinfo{year}{2012}).

\bibitem[{\citenamefont{Foffa and Sturani}(2013)}]{FStail}
\bibinfo{author}{\bibfnamefont{S.}~\bibnamefont{Foffa}} \bibnamefont{and}
  \bibinfo{author}{\bibfnamefont{R.}~\bibnamefont{Sturani}},
  \bibinfo{journal}{Phys. Rev. D} \textbf{\bibinfo{volume}{87}},
  \bibinfo{pages}{044056} (\bibinfo{year}{2013}), \eprint{arXiv:1111.5488
  [gr-qc]}.

\bibitem[{\citenamefont{Galley et~al.}(2015)\citenamefont{Galley, Leibovich,
  Porto, and Ross}}]{GLPR16}
\bibinfo{author}{\bibfnamefont{C.}~\bibnamefont{Galley}},
  \bibinfo{author}{\bibfnamefont{A.}~\bibnamefont{Leibovich}},
  \bibinfo{author}{\bibfnamefont{R.}~\bibnamefont{Porto}}, \bibnamefont{and}
  \bibinfo{author}{\bibfnamefont{A.}~\bibnamefont{Ross}}
  (\bibinfo{year}{2015}), \eprint{arXiv:1511.07379}.

\bibitem[{\citenamefont{Arun et~al.}(2008)\citenamefont{Arun, Blanchet, Iyer,
  and Qusailah}}]{ABIQ08tail}
\bibinfo{author}{\bibfnamefont{K.}~\bibnamefont{Arun}},
  \bibinfo{author}{\bibfnamefont{L.}~\bibnamefont{Blanchet}},
  \bibinfo{author}{\bibfnamefont{B.~R.} \bibnamefont{Iyer}}, \bibnamefont{and}
  \bibinfo{author}{\bibfnamefont{M.~S.} \bibnamefont{Qusailah}},
  \bibinfo{journal}{Phys. Rev. D} \textbf{\bibinfo{volume}{77}},
  \bibinfo{pages}{064034} (\bibinfo{year}{2008}), \eprint{arXiv:0711.0250
  [gr-qc]}.

\bibitem[{\citenamefont{Detweiler}(2008)}]{Det08}
\bibinfo{author}{\bibfnamefont{S.}~\bibnamefont{Detweiler}},
  \bibinfo{journal}{Phys. Rev. D} \textbf{\bibinfo{volume}{77}},
  \bibinfo{pages}{124026} (\bibinfo{year}{2008}), \eprint{arXiv:0804.3529
  [gr-qc]}.

\bibitem[{\citenamefont{Barack and Sago}(2011)}]{BarackS11}
\bibinfo{author}{\bibfnamefont{L.}~\bibnamefont{Barack}} \bibnamefont{and}
  \bibinfo{author}{\bibfnamefont{N.}~\bibnamefont{Sago}},
  \bibinfo{journal}{Phys. Rev. D} \textbf{\bibinfo{volume}{83}},
  \bibinfo{pages}{084023} (\bibinfo{year}{2011}), \eprint{arXiv:1101.3331
  [gr-qc]}.

\bibitem[{\citenamefont{Damour and Sch{\"a}fer}(1988)}]{DS88}
\bibinfo{author}{\bibfnamefont{T.}~\bibnamefont{Damour}} \bibnamefont{and}
  \bibinfo{author}{\bibfnamefont{G.}~\bibnamefont{Sch{\"a}fer}},
  \bibinfo{journal}{Nuovo Cim.} \textbf{\bibinfo{volume}{B101}},
  \bibinfo{pages}{127} (\bibinfo{year}{1988}).

\bibitem[{\citenamefont{Sommerfeld}(1969)}]{Sommerfeld}
\bibinfo{author}{\bibfnamefont{A.}~\bibnamefont{Sommerfeld}},
  \emph{\bibinfo{title}{Atombau und Spektrallinien}}, vol.~\bibinfo{volume}{1}
  (\bibinfo{publisher}{Vieweg}, \bibinfo{address}{Braunschweig},
  \bibinfo{year}{1969}).

\bibitem[{\citenamefont{Mart\'in-Garc\'ia et~al.}(GPL
  2002--2012)\citenamefont{Mart\'in-Garc\'ia, Garc\'ia-Parrado, Stecchina,
  Wardell, Pitrou, Brizuela, Yllanes, Faye, Stein, Portugal et~al.}}]{xtensor}
\bibinfo{author}{\bibfnamefont{J.~M.} \bibnamefont{Mart\'in-Garc\'ia}},
  \bibinfo{author}{\bibfnamefont{A.}~\bibnamefont{Garc\'ia-Parrado}},
  \bibinfo{author}{\bibfnamefont{A.}~\bibnamefont{Stecchina}},
  \bibinfo{author}{\bibfnamefont{B.}~\bibnamefont{Wardell}},
  \bibinfo{author}{\bibfnamefont{C.}~\bibnamefont{Pitrou}},
  \bibinfo{author}{\bibfnamefont{D.}~\bibnamefont{Brizuela}},
  \bibinfo{author}{\bibfnamefont{D.}~\bibnamefont{Yllanes}},
  \bibinfo{author}{\bibfnamefont{G.}~\bibnamefont{Faye}},
  \bibinfo{author}{\bibfnamefont{L.}~\bibnamefont{Stein}},
  \bibinfo{author}{\bibfnamefont{R.}~\bibnamefont{Portugal}},
  \bibnamefont{et~al.}, \emph{\bibinfo{title}{{xAct}: Efficient tensor computer
  algebra for {Mathematica}}} (\bibinfo{year}{GPL 2002--2012}),
  \bibinfo{note}{http://www.xact.es/}.

\bibitem[{\citenamefont{Marchand et~al.}(2016)\citenamefont{Marchand, Blanchet,
  and Faye}}]{MBF16}
\bibinfo{author}{\bibfnamefont{T.}~\bibnamefont{Marchand}},
  \bibinfo{author}{\bibfnamefont{L.}~\bibnamefont{Blanchet}}, \bibnamefont{and}
  \bibinfo{author}{\bibfnamefont{G.}~\bibnamefont{Faye}},
  \bibinfo{journal}{Class. Quant. Grav.} \textbf{\bibinfo{volume}{33}},
  \bibinfo{pages}{244003} (\bibinfo{year}{2016}), \eprint{arXiv:1607.07601
  [gr-qc]}.

\bibitem[{\citenamefont{Gradshteyn and Ryzhik}(1980)}]{GR}
\bibinfo{author}{\bibfnamefont{I.}~\bibnamefont{Gradshteyn}} \bibnamefont{and}
  \bibinfo{author}{\bibfnamefont{I.}~\bibnamefont{Ryzhik}},
  \emph{\bibinfo{title}{Table of Integrals, Series and Products}}
  (\bibinfo{publisher}{Academic Press}, \bibinfo{year}{1980}).

\end{thebibliography}

\end{document}